\documentclass[11pt,eqsecnum
]{article}
\usepackage{amsmath,amsfonts,amssymb,latexsym,theorem,mathrsfs,latexsym,cite,setspace,
comment,color,multirow,ulem,graphicx}
\pdfoutput=1

\makeatletter

\@addtoreset{equation}{section}
\makeatletter

\newcommand{\mas}[1]{\mbox{$\mathscr{#1}$}}

\newcommand{\D}{{\rm d}}
\newcommand{\ti}{\tilde}

\begin{document}

\begin{titlepage}

\begin{flushright}
{
\small YITP-20-127\\
\today
}
\end{flushright}
\vspace{1cm}

\begin{center}
{\LARGE \bf
\begin{spacing}{1}
Static spacetimes haunted by a phantom scalar field: 
\\ \vspace{0.2cm}
classification and global structure in the massless case
\end{spacing}
}
\end{center}
\vspace{.5cm}

\begin{center}
{\large \bf
Cristi{\'a}n Mart\'{\i}nez$^{a}$ and Masato Nozawa${}^{b}$ 
} \\

\vskip 1cm
{\it
${}^a$Centro de Estudios Cient\'{\i}ficos (CECs), Av. Arturo Prat 514, Valdivia, Chile. \\
${}^b$Center for Gravitational Physics, Yukawa Institute for Theoretical Physics, Kyoto University, Kyoto 606-8502, Japan.}\\

\texttt{martinez@cecs.cl, masato.nozawa@yukawa.kyoto-u.ac.jp}

\end{center}

\vspace{.5cm}

\begin{abstract}
We discuss various novel features of $n(\ge 4)$-dimensional spacetimes sourced by a massless (non-)phantom scalar field in general relativity. Assuming that the metric is a warped product of static two-dimensional Lorentzian spacetime and an $(n-2)$-dimensional Einstein space $K^{n-2}$ with curvature $k=0, \pm 1$, and that the scalar field depends only on the radial variable, we present a complete classification of static solutions for both signs of kinetic term. Contrary to the case with a non-phantom scalar field, the Fisher solution is not unique, and there exist two additional metrics corresponding to the generalizations of the Ellis-Gibbons solution and the Ellis-Bronnikov solution. We explore the maximal extension of these solutions in detail by the analysis of null/spacelike geodesics and singularity. For the phantom Fisher and Ellis-Gibbons solutions, we find that there inevitably appear parallelly propagated (p.p) curvature singularities in the parameter region where there are no scalar curvature singularities. Interestingly, the areal radius blows up at these p.p curvature singularities, which are nevertheless accessible within a finite affine time along the radial null geodesics. It follows that only the Ellis-Bronnikov solution describes a regular wormhole in the two-sided asymptotically flat spacetime. 
Using the general transformation relating the Einstein and Jordan frames, we also present a complete classification of solutions with the same symmetry coupled to a conformal scalar field. Additionally, by solving the field equations in the Jordan frame, we prove that this classification is genuinely complete.
\end{abstract}

\vspace{.5cm}

\setcounter{footnote}{0}

\end{titlepage}

\setcounter{tocdepth}{2}
\tableofcontents

\newpage 
%

\section{Introduction}

Scalar fields with negative kinetic energy could constitute a definite menace
to the stability of vacuum states. Still, it seems to be sort of premature to
discard theories with these phantom fields, since some observational data 
for the present day accelerating universe does not exclude the phantom-like equations of state \cite{Aghanim:2018eyx}. 
The viability of phantom fields has been discussed by many authors from classical and quantum
points of view \cite{Caldwell:1999ew,Caldwell:2003vq,Dabrowski:2003jm}.

A discriminative feature of phantom fields is that they fail to obey energy conditions \cite{Maeda:2018hqu}. Energy conditions bear relevance to the positivity of local energy density of matter fields and the causality of energy flux \cite{Hawking:1973uf}. This property puts a strong restriction to the spacetime curvature through Einstein's field equations. A myriad of significant milestones in general relativity--including
a series of singularity theorems \cite{Penrose:1964wq,Hawking:1969sw}, 
positive mass theorems \cite{Schon:1979rg,SchonYau,Witten:1981mf}, area and topology theorem of black holes \cite{Hawking:1971vc}--have postulated the validity of energy conditions. Among these, an important consequence of the null energy condition together with asymptotic flatness is the theorem of topological censorship~\cite{Friedman:1993ty}, which claims that any causal curves starting from past null infinity and ending at future null infinity are homotopically deformable to a trivial curve in the asymptotic region. Specifically, the unrealizability of traversable wormholes is a direct corollary of topological censorship \cite{Galloway}.

Wormholes are eminent bridge structures of spacetime which theoretically provide a way to do interstellar and time travels, and drive warping into another universe. A traversable wormhole was first discussed by the seminal work of Morris and Thorne in 1998 \cite{Morris:1988cz,Morris:1988tu}. Their construction was based upon the static solution sourced by  a  massless phantom scalar field, which has been later identified with the solution already discovered by Ellis \cite{Ellis1973} and Bronnikov \cite{Bronnikov1973}. Although traversable wormholes have been a long-standing arena for paradoxes like closed timelike curves, 
they have attracted resurgence of attention recently, in the context of quantum entanglement and 
teleportation in gauge/gravity duality \cite{Maldacena:2004rf,Gao:2016bin,Maldacena:2017axo,Maldacena:2018lmt}. 
The wormhole throat in this context is supported by quantum matter fields. 
This dual role of wormholes prompts the renewed interest in classical geometry of wormholes, as well as
 the role of classical matter fields violating energy conditions.  In this paper, we shall employ the phantom scalar field as the simplest source to defy energy conditions. A comprehensive study in this simple setting will be likely to witness further progress in the quantum entanglement.

 A static and spherically symmetric {four-dimensional} solution to Einstein's equations with a non-phantom scalar field is 
 unique and  constitutes a {three}-parameter family of metrics. This exact solution was introduced by 
  Fisher \cite{Fisher:1948yn} in 1948, and rediscovered in successive decades by  Bergmann and Leipnik \cite{Bergmann:1957zza}, Buchdahl \cite{Buchdahl:1959nk},  Janis, Newman and Winicour \cite{jnw1968}, Ellis \cite{Ellis1973}, Bronnikov \cite{Bronnikov1973} and Wyman \cite{Wyman:1981bd}.\footnote{The formulae presented in \cite{Bergmann:1957zza},\cite{Bronnikov1973} and \cite{Wyman:1981bd} also include the case of a phantom scalar field.} In this article,  we refer to this configuration as the Fisher solution, as well as the higher dimensional and topological generalizations. A complexification $\phi \to i \phi$ of the scalar field in the Fisher solution 
then trivially solves Einstein's equations with a phantom scalar field. In the phantom case,  the phantom Fisher solution is {\it not} unique in the static and spherically symmetric system, and  there exist two additional offbeat solutions, as first pointed out by Ellis \cite{Ellis1973} in four dimensions.\footnote{In \cite{Bergmann:1957zza}, the authors do not provide an explicit treatment for the phantom field and properly address only the Fisher class ~\ref{Fisher class}.  As far as we know, the first analysis of solutions with a phantom scalar field was given by H. G. Ellis in 1973 \cite{Ellis1973}.} One is a renowned wormhole and the other is a solution later rediscovered and generalized by Gibbons \cite{Gibbons:2003yj,Gibbons:2017jzk}. This nonuniqueness feature is in marked contrast to the non-phantom case. 
However, very little is known for the global structures of each solution, except for the 
Ellis-Bronnikov wormhole solution with a vanishing mass. 
In order to see what kind of peculiarity comes out when the energy conditions are 
false, the global spacetime structure is a central subject to be studied. 
To fill this gap is one of the aims of the current paper. 
 
We generalize the  illustrious work of Ellis \cite{Ellis1973} and 
inaugurate the classification of static solutions in Einstein-(phantom-)scalar system in arbitrary dimensions. 
We extend the setup of \cite{Ellis1973} in such a way that the $n$-dimensional metric is a warped spacetime of static two-dimensional Lorentzian spacetime and an $(n-2)$-dimensional Einstein space $K^{n-2}$ with curvature $k=0, \pm 1$, and that the scalar field depends only on the radial variable. For each sign of the curvature of $K^{n-2}$, we perform the exhaustive classification of solutions and find their explicit expressions in a closed form. 
Restricting to the spherically symmetric case, we next examine the global structure of the solutions
in detail for all range of the solution parameters. We verify that the Fisher solution and the Ellis-Gibbons solution 
admit a parameter range under which there arise no scalar curvature singularities. 
Some preceding works have therefore concluded that these solutions describe wormholes. 
We demonstrate, however, that there exist spacetime points where some Riemann tensor components in a frame which is parallelly propagated along the radial null geodesics diverge, i.e., these solutions possess naked p.p curvature singularities. This crucial property has been overlooked in the literature. These solutions therefore fail to describe regular wormhole solutions and the only regular wormholes in this class are 
the Ellis-Bronnikov solution.  This is one of the main results in this paper. 
Next, we develop {two procedures} which transform the system with a massless scalar field {(Einstein frame)} to the system with a conformally coupled scalar field {(Jordan frame)} for both signs of the kinetic term. Exploiting this framework, 
we perform a complete classification of solutions with a conformally coupled scalar field admitting the same symmetry as the original solutions. { In addition,  as a consistency check, we  find the same classification by solving directly the field equations in the Jordan frame.}

We organize the present paper as follows: 
In the next section, we classify static solutions with a  {massless} scalar field and 
derive three distinct family of solutions. In section \ref{properties}, 
we elucidate physical properties of these solutions in detail. In section  \ref{conformal} we present the classification for a {massless} conformally {coupled} scalar field. We conclude our paper in section \ref{sec:conclusion} with some future prospects. 
Useful curvature formulae are encapsulated in Appendix \ref{app:geom}.

Our basic notations follow~\cite{wald}.
The conventions of curvature tensors are 
$[\nabla _\rho ,\nabla_\sigma]V^\mu ={R^\mu }_{\nu\rho\sigma}V^\nu$ 
and ${R}_{\mu \nu }={R^\rho }_{\mu \rho \nu }$.
The Lorentzian metric is taken to be the mostly plus sign, and 
Greek indices run over all spacetime indices. 
The $n$-dimensional gravitational
constant is denoted by  $\kappa_{n}=8\pi G_n$.

\section{Classification}
\label{sec:classification}

{The first part of this} paper focuses on the  $n(\ge 4)$-dimensional spacetimes involving a massless scalar field {in the Einstein frame} described by the action
\begin{align}
\label{action}
S[g_{\mu\nu},\phi]=&\int \D^nx\sqrt{-g}\biggl(\frac{1}{2\kappa_{n}}{R}-\frac12{\epsilon}({\nabla} \phi)^2 \biggl),
\end{align}
where $\epsilon=+1$ for a conventional scalar field and $\epsilon=-1$ for a phantom field, with $\kappa_n >0$ denoting the gravitational constant. The Einstein and scalar field equations following from the above action read 
\begin{align}
E_{\mu\nu}:=&{R}_{\mu\nu}-\frac12g_{\mu\nu}{R}-\kappa_{n}T_{\mu\nu}=0,
\label{em-kg}\\
\square \phi=&0, \label{KGeq}
\end{align}
respectively, where the energy-momentum tensor for the massless scalar field is
\begin{align}
T_{\mu\nu}: =&\epsilon \left((\nabla_\mu \phi)(\nabla_\nu \phi)-\frac12 g_{\mu\nu} (\nabla\phi)^2\right)
.\label{Tab-scalar}
\end{align}
For $\epsilon=1$, the scalar field satisfies all the standard energy conditions \cite{Maeda:2018hqu}.
For $\epsilon=-1$, the phantom field enjoys the negative kinetic energy 
and therefore mediates a repulsive force. 

We consider the $n$-dimensional warped product metric of an $(n-2)$-dimensional constant curvature space $(K^{n-2}, \gamma _{ij})$
and a two-dimensional spacetime $(M^2, g_{AB})$. Here $\gamma_{ij}(z)$ is the metric on the $(n-2)$-dimensional Einstein space $K^{n-2}$, whose Ricci tensor is given by ${}^{(\gamma)}{R}_{ij}=k(n-3)\gamma_{ij}$, where $k =1,0,-1$. 
This $2+(n-2)$ dimensional curvature decomposition has been intensively studied in the literature, which we summarize in appendix \ref{app:geom} to make this paper self-contained. 
Along this article, we further focus on the static solutions. Choosing the coordinates of $M^2$ as $t$ and $x$, we shall consider solutions described by the following class of metric
\begin{align}
\D s^2=&-F(x)^{-2}\D t^2+F(x)^{2/(n-3)}G(x)^{-(n-4)/(n-3)}\biggl(\D x^2+G(x)\gamma_{ij}(z)\D z^i\D z^j\biggl).\label{gauge-higher}
\end{align}
Since the metric is static, the vector field $\partial/\partial t$ denotes its timelike Killing vector.  The above choice of the coordinates for describing static configurations was proven to be very convenient to perform a classification in the presence of an additional electric field \cite{Maeda:2016ddh}. 

We further assume that the scalar field $\phi$ does not depend both on time $t$ 
and the coordinates $z^i$ describing $K^{n-2}$, namely, $\phi$ is a scalar function of $x$. As shown below, for the case where the areal radius $S:=(F^2 G)^{1/[2(n-3)]}$ is constant, we find no solutions. In consequence,  $x$  is considered as the ``radial'' coordinate conjugate to $t$.

Plugging $\phi=\phi(x)$ into (\ref{KGeq}), one can immediately integrate  it once  to find 
\begin{align}
\frac{\D \phi}{\D x}=\frac{\phi_1}{G},
\end{align}
where $\phi_1$ is a constant. 
The combination $E^x{}_{x}+E^i{}_{i}=0$ (no summation over $i$) gives the master equation for $G(x)$:
\begin{align}
\frac{\D ^2G}{\D x^2}-2k(n-3)^2=0. \label{master-G}
\end{align}
Upon using (\ref{master-G}), the rest of Einstein' equations $E_{\mu\nu}=0$ boil down to
\begin{align}
E^t{}_{t}=&F^{\frac{-2}{n-3}} G^{\frac{n-4}{n-3}}\left[\frac{n-2}{8(n-3)}\biggl\{8F^{-1}\frac{\D ^2F}{\D x^2}-4F^{-2}\biggl(\frac{\D F}{\D x}\biggl)^2+8F^{-1}G^{-1}\frac{\D F}{\D x}\frac{\D G}{\D x} \right. \nonumber \\
& \qquad \qquad \qquad 
\left. -G^{-2}\biggl(\frac{\D G}{\D x}\biggl)^2 +4k(n-3)^2G^{-1}\biggl\}+\kappa_{n}{\epsilon}\frac{\phi_1^2}{2}G^{-2}\right],\label{constraint} \\
E^x{}_{x}=&F^{\frac{-2}{n-3}} G^{\frac{n-4}{n-3}}\left[\frac{n-2}{8(n-3)}\biggl\{-4F^{-2}\biggl(\frac{\D F}{\D x}\biggl)^2+G^{-2}\biggl(\frac{\D G}{\D x}\biggl)^2 -4k(n-3)^2G^{-1}\biggl\}-\kappa_{n}{\epsilon}\frac{\phi_1^2}{2} G^{-2}\right],
\end{align}
which are further simplified to 
\begin{align}
F^{-1}\frac{\D ^2F}{\D x^2}-F^{-2}\biggl(\frac{\D F}{\D x}\biggl)^2+F^{-1}G^{-1}\frac{\D F}{\D x}\frac{\D G}{\D x}&=0,\label{master-F2}\\
F^{-2}\biggl(\frac{\D F}{\D x}\biggl)^2-\frac14G^{-2}\biggl(\frac{\D G}{\D x}\biggl)^2+k(n-3)^2G^{-1}+
{\epsilon}\frac{(n-3)\kappa_{n} \phi_1^2}{n-2}G^{-2}&=0. \label{master-F1}
\end{align}

We are now ready to perform the classification of solutions.
$G(x)$ and $F(x)$ are obtained from Eqs.~(\ref{master-G}) and (\ref{master-F2}), respectively. 
The final equation (\ref{master-F1}) corresponds to a constraint on them,  giving rise to the 
scalar field configuration.
The general solution to (\ref{master-G}) is 
\begin{align}
G(x)=k(n-3)^2x^2+G_1x+G_0,\label{G-sol-gen}
\end{align}
where $G_0$ and $G_1$ are constants.
The following analysis divides into subclasses 
depending how many different real roots the function $G(x)$ admits.

\subsection{Solutions for $k=\pm 1$}

First let us consider the case of $k=1,-1$.
In this case, there are three sub-cases: (i) $G(x)$ has two real roots, (ii) $G(x)$ has one degenerate real root, and (iii) $G(x)$ has no real root.

\subsubsection{Fisher class: $G(x)$ has two distinct real roots} \label{Fisher class}

In this case, we write $G(x)$ as
\begin{align}
G(x)=k(n-3)^2(x-a)(x-b).
\end{align}
with new parameters $a$ and $b(\ne a)$.
Then, the general solution of the master equation (\ref{master-F2}) is
\begin{align}
F(x)=A\biggl({\sigma}\frac{x-a}{x-b}\biggl)^{-\alpha/2},
\end{align}
where $\alpha$ and $A$ are constants. We have introduced an ancillary constant  
${\sigma}:={\rm sgn}[(x-a)/(x-b)]=\pm 1$
for convenience.
The constraint (\ref{master-F1}) gives
\begin{align}
\phi_1^2={\epsilon}\frac{k^2(n-2)(n-3)^3(1-\alpha^2)(a-b)^2}{4\kappa_n}.
\end{align}
The reality of $\phi$ asks for $\epsilon (1-\alpha^2)\ge 0$. 
The solution is therefore given by 
\begin{align} \label{Fisherk}
F(x)=&A\biggl({\sigma}\frac{x-a}{x-b}\biggl)^{-\alpha/2},\qquad G(x)=k(n-3)^2(x-a)(x-b),\\
\phi=&\phi_0\pm \sqrt{{\epsilon}\frac{(n-2)(1-\alpha^2)}{4(n-3)\kappa_n}}\ln\biggl({\sigma}\frac{x-a}{x-b}\biggl),
\end{align}
where $\phi_0$ is yet another integration constant. 
Here, the Lorentz signature of the metric demands $G(x)> 0$, which amounts to
${\sigma} k=1$. 

Let us find a more friendly expression of this solution.
By the coordinate transformations $t=A{\bar t}$ and
 $x-b=r^{n-3}/[A(n-3)]$, the solution is recast into 
\begin{align}
\label{JNWk}
\D s^2=&-f(r)^{\alpha}\D {\bar t}^2+ f(r)^{-(\alpha+n-4)/(n-3)}\biggl(\D r^2+r^2f(r)\gamma_{ij}(z)\D z^i\D z^j\biggl),\\
\phi=&{\phi}_0\pm \sqrt{{\epsilon}\frac{(n-2)(1-\alpha^2)}{4(n-3)\kappa_n}}\ln {f(r)},\qquad f(r)={\left|k-\frac{M}{r^{n-3}}\right|},
\end{align}
where we have renamed the parameter as $M={\sigma} A(n-3)(a-b)$. 
The scalar field becomes phantom ($\epsilon=-1$) for $\alpha^2>1$ and conventional if $\alpha^2 \le 1$. 
For a given curvature constant  $k$ of $K^{n-2}$, this is a {three}-parameter family of solutions characterized by $M, \alpha$ and $\phi_0$. The  
spherically symmetric case ($k=1$) in $n=4$ with a conventional scalar field
corresponds to the Fisher solution found in \cite{Fisher:1948yn,Bergmann:1957zza,Buchdahl:1959nk,
jnw1968,Ellis1973,Bronnikov1973,Wyman:1981bd}. For $n>4$, $\epsilon=1$ and $K^{n-2}$ chosen as the unit ($n-2$)-dimensional round sphere, the solution \eqref{JNWk} matches the Xanthopoulos -Zannias solution \cite{JNWhigher}, which was comprehensively analyzed in \cite{Abdolrahimi:2009dc}.

\subsubsection{Ellis-Gibbons class: $G(x)$ has one real degenerate root}

Let us consider the case in which $G(x)$ admits a two-fold real root.  
In this case, we write $G(x)$ as
\begin{align}
G(x)=k(n-3)^2(x-a)^2.
\end{align}
Then, the general solution of the master equation (\ref{master-F2}) is
\begin{align}
F(x)=F_0e^{x_0/(x-a)},
\end{align}
where $F_0$ and $x_0$ are  constants.
The constraint (\ref{master-F1}) gives
\begin{align}
\phi_1^2=-{\epsilon}\frac{k^2(n-2)(n-3)^3x_0^2}{\kappa_n}.
\end{align}
To render the scalar field real, we are forced to choose $\epsilon=-1$, 
which allows one to make the right hand side of this equation positive. 
Hence the solution exists only for a phantom field and is given by 
\begin{align} \label{Ellis-Gibbonsk}
F(x)=&F_0e^{x_0/(x-a)},\qquad G(x)=k(n-3)^2(x-a)^2, \qquad 
\phi=\phi_0\pm \sqrt{-\epsilon\frac{(n-2)x_0^2}{(n-3)\kappa_n}}\frac{1}{x-a}.
\end{align}
Obviously, the Lorentz signature $G(x)>0$ is assured only for $k=1$. 
Since the metric is written in an awkward form, let us change 
the coordinates by $t=F_0{\bar t}$ and $x-a=r^{n-3}/[(n-3)F_0]$, 
which bring the solution into a  more  familiar form\footnote{As far as the authors know, 
the solution has first appeared in the literature 
in \cite{Yilmaz}, where an alternative theory for gravity was proposed. There, an analogue of the Ricci tensor was defined with a minus sign, turning in practice the scalar field into a phantom one.
}
\begin{align}
\label{Gibbonsol}
\D s^2=&-e^{-M/r^{n-3}}\D {\bar t}^2+e^{M/[(n-3)r^{n-3}]}\biggl(\D r^2+r^2\D \Sigma_{k=1,n-2}^2\biggl),\\
\phi=&\phi_0\pm \sqrt{-{\epsilon}\frac{n-2}{4(n-3)\kappa_n}}\frac{M}{r^{n-3}},
\end{align}
where we have set $M:=2(n-3)F_0x_0$.
This is a two-parameter, given by the constants $M$ and $\phi_0$, family of phantom-scalar solutions and sometimes referred to as the ``exponential metric.''  When $n=4$ and $\D\Sigma^2_{k=1,2}$ being a metric of round sphere, this solution reduces to the one discovered by Ellis \cite{Ellis1973} and Bronnikov \cite{Bronnikov1973}, which was generalized by Gibbons \cite{Gibbons:2003yj,Gibbons:2017jzk} as explained in later section \ref{short summary}.

\subsubsection{Ellis-Bronnikov class: $G(x)$ has no real roots} \label{Ellis-Bronnikov class}

Lastly, we investigate the case in which $G(x)$ admits no real roots. 
In this case, we write $G(x)$ as
\begin{align}
G(x)=k(n-3)^2x^2+G_0, 
\end{align}
where we have used the degree of freedom to change the origin of $x$. 
For the absence of real roots, $k $ and $G_0$ must carry the same sign. 
Then a unique option in this case is $k=1$ and $G_0>0$: otherwise 
one cannot keep the Lorentzian signature of the metric. 
{Thus}, the general solution of the master equation (\ref{master-F2}) is\footnote{\label{FNarctan}By virtue of $\arctan (x)+\arctan(1/x)=\pi/2$ for $x>0$, the function $\arctan (x)$ has been used extensively for the analysis of Ellis-Bronnikov solution. Here we decided to choose $\arctan (1/x)$, which makes the asymptotic analysis easier. In contrast, the argument of spacetime extension is 
simpler if we make the choice $\arctan (x)$, as will be discussed in section~\ref{sec:EB}.}
\begin{align}
F(x)=F_0\exp\left(\beta\arctan\biggl(\frac{x_0}{x}\biggl)\right),
\end{align}
where $F_0$ and $\beta$ are  integration constants and we have defined
$G_0=(n-3)^2x_0^2$.
The constraint (\ref{master-F1}) gives
\begin{align}
\phi_1^2=-{\epsilon} x_0^2\frac{(n-2)(n-3)^3(1+\beta^2)}{\kappa_n}.
\end{align}
Again, only the phantom case $\epsilon=-1$ is allowed and $\phi$ is found to be
\begin{align}
\phi={\phi}_0\pm\sqrt{-{\epsilon}\frac{(n-2)(1+\beta^2)}{\kappa_n(n-3)}}\arctan\biggl(\frac{x_0}
{x}\biggl).
\end{align}
With this form of metric at hand, 
the coordinate change $t = F_0 \bar t$, $x=2x_0 r^{n-3}/M$ with $M=2(n-3)x_0F_0$
allows us to have a more suggestive form
\begin{align}
\label{Ellis-Bronnikov}
\D s^2&=-e ^{-2\beta U(r)}\D \bar t^2+e^{2\beta U(r)/(n-3)}V(r)^{1/(n-3)}
\left(\frac{\D r^2}{V(r)}+r^2 \D \Sigma_{k=1,n-2}^2 \right),\\
\phi&=\phi_0\pm\sqrt{-{\epsilon}\frac{(n-2)(1+\beta^2)}{\kappa_n(n-3)}} U(r),
\\
U(r):&= \arctan \left(\frac{M}{2r^{n-3}}\right), \qquad 
V(r):= 1+\frac{M^2}{4r^{2(n-3)}}.
\label{Ellis-BronnikovUV}
\end{align}
This is a three-parameter family of solutions and reduces to the Ellis-Bronnikov phantom-scalar solution in $n=4$ \cite{Ellis1973,Bronnikov1973}.
A higher dimensional solution with $\beta=0$ was first derived in \cite{Torii:2013xba}, but the 
solution is expressed in an implicit fashion,  
since the authors employed a gauge $g_{rr} g_{tt}=-1$.

\subsection{Solutions for $k=0$}

Let us next consider the case of $k=0$, namely, the one where $K^{n-2}$ is chosen to be a Ricci-flat Riemannian space. 
In this case, Eq.~(\ref{master-G}) is integrated to give
\begin{align}
G(x)=G_1x+G_0, \label{sol-G3}
\end{align}
where $G_0$ and $G_1$ are constants.
Upon integration, the scalar field is given by 
\begin{align}
\phi(x)=\phi_0+\frac{\phi_1}{G_1}\ln |G_1x+G_0 |  \label{sol-phi4}
\end{align}
for $G_1\ne 0$ and 
\begin{align}
\phi(x)=\phi_0+\frac{\phi_1}{G_0}x \label{sol-phi5}
\end{align}
for $G_1=0$.
We will study these two sub-cases separately.

\subsubsection{Fisher class: $G_1\ne 0$}

In this case, the general solution of the master equation (\ref{master-F2}) is
\begin{align}
F(x)=&A(G_1x+G_0)^{\alpha/2}, \label{sol-F6}
\end{align}
where $A$ and $\alpha$ are constants.
The constraint (\ref{master-F1}) gives
\begin{align}
\phi_1^2=&\epsilon\frac{(n-2)(1-\alpha^2)G_1^2}{4(n-3)\kappa_{n}},
 \label{sol-real-cond1} 
\end{align}
requiring $\epsilon (1-\alpha^2)>0$. 
The solution is therefore given by 
\begin{align}
F(x)=&A(G_1x+G_0)^{\alpha/2},\qquad G(x)=G_1x+G_0, \label{k=0 case 1}\\
\phi(x)=&\phi_0\pm\sqrt{\epsilon\frac{(n-2)(1-\alpha^2)}{4(n-3)\kappa_{n}}}\ln|G_1x+G_0|.\label{Type-V-phi}
\end{align}
By the coordinate transformations $t=AG_1^{\alpha}{\bar t}/(n-3)^{\alpha}$ 
and $G_1x+G_0=[(n-3)^{\alpha-1}/(AG_1^{\alpha-1})]r^{n-3}$ 
$M=-AG_1^{\alpha+1}/(n-3)^{\alpha+1}$, 
a straightforward computation shows that one can transform the above metric 
into the $k=0$ Fisher solution (\ref{JNWk}) up to 
a redefinition of the constant $\phi_0$. 
For $n=4$ and the conventional scalar field, this solution was first found in \cite{Tabensky-Taub, Singh, Buchdahl1978} and has been also discussed in \cite{Vuille:2007ws}. 

\subsubsection{$G_1=0$}

In this case, the Lorentzian signature requests $G_0>0$.
The general solution of the master equation (\ref{master-F2}) is
\begin{align}
F(x)=&F_0e^{x/x_0},\label{sol-F8}
\end{align}
where $F_0$ and $x_0$ are constants.
The constraint (\ref{master-F1}) gives
\begin{align}
\phi_1^2=-\epsilon\frac{(n-2)G_0^2}{(n-3)\kappa_nx_0^2}. \label{sol-rel8}
\end{align}
This class of solutions is therefore allowed only for a phantom case ($\epsilon=-1$). 
Hence, by Eq.~(\ref{sol-real-cond1}), the solution is given by 
\begin{align}
F(x)=F_0e^{x/x_0},\qquad G(x)=G_0,\qquad
 \phi(x)=\phi_0\pm\sqrt{-\epsilon\frac{n-2}{(n-3)\kappa_nx_0^2}}  x.
\end{align}
By the coordinate transformations 
$t=\sqrt{G_0} \, {\bar t}$ and  $x= \sqrt{G_0} {\bar x} -x_0\ln(F_0\sqrt{G_0})$, the solution becomes 
\begin{align}
\label{Buchdahl}
&\D s^2=-e^{-2\zeta{\bar x}}\D {\bar t}^2+e^{2\zeta{\bar x}/(n-3)}\biggl(\D {\bar x}^2+\D {\Sigma}_{k=0, n-2}^2\biggl),\\
&\phi(\bar x)=\phi_0\pm\sqrt{-\epsilon\frac{(n-2)\zeta^2}{(n-3)\kappa_n}}{\bar x},
\end{align}
where we have introduced a new parameter
$\zeta :=\sqrt{G_0}/x_0$ and redefined $\phi_0$. 
This is a {two}-parameter family of phantom-scalar solutions. The early papers dealing with planar symmetric solution in four dimensions do not show this solution since the coordinates chosen are not suitable for this case. It was in the appendix of \cite{Erices:2015xua} that provides this solution first time ever using other radial coordinate.\footnote{At (A38) in the appendix of \cite{Erices:2015xua}, set $\sigma_1=0$, $\sqrt{a}=-2 u_1$ and  change the radial coordinate as $u_0+u_1 r= e^{\zeta \bar{x}}$.} The extension to higher dimensions
seems to be original.

\subsection{Short summary and remarks} \label{short summary}

We have classified static solutions in Einstein-scalar system with 
either sign of kinetic terms. It has not been fully recognized that the richness 
of the plausible solutions depends sensitively to the sign of the scalar kinetic term. 
We have obtained a complete catalogue of solutions with four distinct family: 
the generalized Fisher solution (\ref{JNWk}) (valid for $k=0, \pm 1$), 
the generalized Ellis-Gibbons solution  (\ref{Gibbonsol}), 
the generalized Ellis-Bronnikov solution  (\ref{Ellis-Bronnikov}) 
and a new plane-symmetric  solution (\ref{Buchdahl}). 
We incorporated ``generalized'' to highlight that the present solution is more general 
than four dimensional counterparts in that $\D \Sigma_{k,n-2}^2$ is not necessarily a maximally symmetric space. Aside from the Fisher class, all other solutions are permissible only for the phantom case. 
In other words, the non-phantom solutions covered by the metric ansatz 
 (\ref{gauge-higher}) are exhausted by the Fisher solution,  
which has been broadly discussed in the literature. 
In the phantom case, on the other hand, the class of metric  (\ref{gauge-higher})
encompasses a wide variety of solutions, as we have argued. 
In four dimensions, the Fisher solution, the Ellis-Gibbons solution and the 
Ellis-Bronnikov solution are related to each other by complexification of parameters and
infinite boost limit \cite{Gibbons:2017jzk}.

\subsubsection{Gibbons solution}

Let us  go into the structure of 
the generalized Ellis-Gibbons solution (\ref{Gibbonsol}). 
An innovative feature of Ellis-Gibbons solution ($n=4$) is that one can superpose 
the point sources  in the exponent of the metric components 
without further restrictions \cite{Gibbons:2003yj}, in that the term $M/r$ can be 
replaced by the arbitrary harmonics $\sum_i M_i/|{\bf x}-{\bf x}_i|$ in flat space $\mathbb R^3$. 
In the present setting, we can extend this into a more general form \cite{Maeda:2019tqs}
\begin{align}
\label{multiGibbons}
\D s^2 =- e^{{-H}} \D t^2+e^{{H/(n-3)}} h_{IJ}\D x^I \D x^J, \qquad 
\phi=\pm \sqrt{-{\epsilon}\frac{n-2}{4(n-3)\kappa_n}}H, \qquad 
\Delta_{h} H=0,
\end{align}
where $h_{IJ}$ is the arbitrary ($n-1$)-dimensional Ricci flat metric ${}^{(h)}R_{IJ}=0$. 
If we take $h_{IJ}$ as a flat metric on the Euclid space and consider the point source  at the origin $H=M/r^{n-3}$,
the above metric reduces to  the generalized Ellis-Gibbons solution (\ref{Gibbonsol}).
It is worthwhile to observe that this metric (\ref{multiGibbons}) also includes 
(\ref{Buchdahl}) if we consider the linear harmonics. To see this explicitly, let us consider the flat space $\mathbb R^{n-1}$ for which 
the general solution to $\Delta_{\mathbb R^{n-1}}H=0$ reads
\begin{align}
\label{}
H_{\mathbb R^{n-1}} = C^{(0)} +C_I ^{(1)}x^I +C^{(2)}_{IJ}x^Ix^J
+\sum_{i} \frac{M_i}{|{\bf x}-{\bf x}_i|^{n-3}},
\end{align}
where $C^{(0)} $, $C_I ^{(1)}$ and $C^{(2)}_{IJ}$ are constants with 
${\rm Tr}\,C^{(2)}=0$, and $x^I_i$ are the loci of distributional sources. The solution (\ref{Buchdahl}) is recovered, provided only the linear term $C_1 ^{(1)}$ is nonvanishing. 
In what follows, we shall refer to the solution (\ref{multiGibbons}) as the Gibbons solution.

The Gibbons solution (\ref{multiGibbons}) has a striking similarity to the Majumdar-Papapetrou solution \cite{Majumdar:1947eu,Papapetrou}, for which 
the electromagnetic repulsive force between charged point sources compensates
the gravitational attraction. This precise cancellation of forces is a primary reason behind
the linearization of gravitational field equations. 
The Gibbons solution (\ref{multiGibbons}) realizes the delicate balance 
by the repelling force induced by a phantom field. 
In spite of the close resemblance, a fundamental difference between these solutions is that 
the Majumdar-Papapetrou solution preserves supersymmetry \cite{Gibbons:1982fy}, whereas the multi-Gibbons solution does not seem to admit any Killing spinors.

\subsubsection{Spacetime with constant scalar invariants}

Let us focus on the metric (\ref{k=0 case 1}), which is the $k=0$ Fisher class. 
Recently, it has been noted\footnote{Indeed, it was shown  in \cite{Erices:2015xua} that CSI spacetimes arise from phantom cylindrical configurations, which contain the planar-symmetric ones as particular cases.} in \cite{Erices:2015xua} that the four-dimensional phantom configuration of the $k=0$ Fisher class with $\alpha =-2$, 
\begin{align} \label{CSI,n=4}
\D s^2=&-(G_1x+G_0)^{2}A^{-2}\D t^2+(G_1x+G_0)^{-2}A^{2}\D x^2+(G_1x+G_0)^{-1}A^{2}(\D y^2+\D z^2), \\
\phi(x)=&\phi_0\pm\sqrt{
\frac{-3\epsilon}{2\kappa_{4}}}\ln|G_1x+G_0|,\label{TCSI-phi}
\end{align}
corresponds to a constant scalar invariant (CSI) spacetime \cite{Coley:2005sq,Coley:2009tx}, namely, a spacetime featuring all polynomial scalar invariants constructed from the Riemann tensor and its covariant derivatives constant. To find the higher dimensional extension, we write down the following curvature invariants for the metric (\ref{k=0 case 1}):
\begin{align} 
\label{R k=0 case 1}
R=&\frac{ (n-2)\left(1-\alpha ^2\right) A^{-\frac{2}{n-3}} G_1^2(G_1x+G_0)^{-\frac{\alpha +n-2}{n-3}}}{4 (n-3)},\\
R_{\mu \nu}R^{\mu \nu}=&\frac{(n-2)^2\left(1-\alpha ^2\right)^2 A^{-\frac{4}{n-3}} G_1^4(G_1x+G_0)^{-\frac{2 (\alpha +n-2)}{n-3}}}{16 (n-3)^2}.
\end{align}
These invariant quantities become constant if $\alpha= 2-n$ for any Ricci-flat space $K^{n-2}$.
By means of the following replacements in (\ref{k=0 case 1}) and (\ref{Type-V-phi}),
\begin{align}
G_1x+G_0 &=A^{2/(n-3)}\ell^2/r^2, \qquad  \ell=2 A^{1/(n-3)}/G_1, \qquad 
\bar t=A^{1/(n-3)}t, \notag \\
 \bar \phi_0&=\phi_0\pm \sqrt{-\frac{\epsilon(n-1)(n-2)}{\kappa_n}} \frac{\ln A}{n-3},
\end{align} 
these special spacetimes, realized only by a phantom scalar field, are conveniently described by the fields
\begin{align}
\label{H-CSI}
\D s^2=- \left(\frac{r}{\ell}\right)^{4-2n} \D \bar t^2+\frac{\ell^2}{r^2}\D r^2+
\frac{r^2}{\ell^2} \D \Sigma_{k=0, n-2}^2, \qquad 
\phi=\bar \phi_0\pm \sqrt{-\frac{\epsilon(n-1)(n-2)}{\kappa_n}}\ln \left(\frac{\ell}{r}\right).
\end{align}
If $K^{n-2}$ is further restricted to the flat space, the invariant $R_{\mu\nu\rho\sigma}R^{\mu\nu\rho\sigma}$ is also constant, and the spacetime \eqref{H-CSI} becomes CSI since it is locally homogeneous. Under this choice of  $K^{n-2}$,  \eqref{H-CSI} is nothing but the Lifshitz  spacetime with a negative  dynamical exponent $z=2-n$ \cite{Kachru:2008yh}.

\section{Properties of spherically symmetric solutions}
\label{properties}

In this section, we focus on the spherically symmetric case 
and study the physical properties of the solutions obtained in the previous section.
In particular, we examine the asymptotic configuration and the nature of the singularity, which 
are used to clarify the  global structure of the spacetime.
We discuss the Fisher class in full detail, since the techniques are directly borrowed from 
section \ref{sec:propJNW}.

\subsection{Fisher class}
\label{sec:propJNW}

The spherically symmetric phantom Fisher solution in $n(\ge 4)$ dimensions is given by 
\begin{align}
\D s^2=&-f(r)^{\alpha}\D { t}^2+ f(r)^{-(\alpha+n-4)/(n-3)}\biggl(\D r^2+r^2f(r)\D \Omega_{n-2}^2 \biggl),
\label{JNWsol}\\
\phi=&{\phi}_0\pm \sqrt{{\epsilon}\frac{(n-2)(1-\alpha^2)}{4(n-3)\kappa_n}}\ln {f(r)},\qquad f(r)=1-\frac{M}{r^{n-3}},  \label{JNWsolphi}
\end{align}
where $\D \Omega_{n-2}^2$ is the standard metric of a unit ($n-2$)-sphere. 
The domain of $r$ in \eqref{JNWsol}  and \eqref{JNWsolphi} is chosen to ensure $f(r)>0$, 
i.e.,  $r_{\rm s}<r<\infty$ for $M>0$, where $r_{\rm s}:=M^{1/(n-3)}$, 
and $0<r<\infty$ for $M\le 0$.
We have omitted the bar from the time coordinate to simplify the notation. 
Vigorous works have been carried out for the higher-dimensional solution with $\epsilon=1$ \cite{JNWhigher,Abdolrahimi:2009dc}.

Defining an isotropic radial coordinate $\rho$ by\footnote{Expressions of the form $x^{2y}$ should be understood as $(x^2)^y$ for any finite real numbers $x$ and $y$.} 
\begin{align}
\label{}
r=\rho \left(1+\frac{M}{4\rho^{n-3}}\right)^{{2}/(n-3)},
\end{align}
the metric can be brought into the following form
\begin{align}
\label{JNWiso}
\D s^2&= - \left(\frac{1-\frac{M}{4\rho^{n-3}}}{1+\frac{M}{4\rho^{n-3}}}\right)^{2\alpha} \D t^2
+\left(1+\frac{M}{4\rho^{n-3}}\right)^{\frac{4}{n-3}} 
\left(\frac{1-\frac{M}{4\rho^{n-3}}}{1+\frac{M}{4\rho^{n-3}}}\right)^{\frac{2(1-\alpha)}{n-3}} 
(\D \rho^2+\rho^2 \D \Omega_{n-2}^2), \\
\phi &=\phi_0\pm \sqrt{{\epsilon}\frac{(n-2)(1-\alpha^2)}{(n-3)\kappa_n}}\ln \left|\frac{1-\frac{M}{4\rho^{n-3}}}{1+\frac{M}{4\rho^{n-3}}}\right|.
\end{align}

\subsubsection{Mass} \label{section-mass}

At infinity ($\rho =\infty$), the metric \eqref{JNWiso}  can be expanded as 
\begin{align}
\label{}
\D s^2\simeq -\left(1-\frac{\alpha M}{\rho^{n-3}}\right)\D t^2+
\left(1+\frac{\alpha M}{(n-3)\rho^{n-3}}\right) (\D \rho^2+\rho^2 \D \Omega_{n-2}^2).
\end{align}
Hence, the spacetime is asymptotically flat. As established in \cite{Arnowitt:1962hi}, for vacuum solutions or those where the matter fields decay fast enough at infinity, the ADM mass can be extracted by comparison with the asymptotic form of the Schwarzschild metric. This is the case for the scalar field with $\phi_0=0$, which behaves as $ \mathcal O(\rho^{3-n})$ at infinity. For this class of configurations, 
a comparison with the asymptotic form of the metric \cite{myersperry1986} 
gives  the ADM mass as
\begin{align}
\label{MADM}
M_{\rm ADM}=\frac{(n-2) \Omega_{n-2} }{2\kappa_n} \alpha M,
\end{align}
where $ \Omega_{n-2} $ is the area of the unit ($n-2$)-sphere
\begin{align}
\Omega_{n-2} :=\frac{2\pi^{(n-1)/2}}{\Gamma((n-1)/2)}.
\label{unitarea}
\end{align}

Although the metric is independent of $\phi_0$, it corresponds to the value of the scalar field at infinity and seems to be of physical relevance. Thus, the choice $\phi_0=0$ is restrictive and unnecessary. A more general and robust formalism for determining the charges associated to asymptotic symmetries was provided by Regge and Teitelboim \cite{Regge:1974zd}. In this formalism the generators are built with suitable boundary terms ensuring they have well-defined functional derivatives. Because one obtains the variation of the generators, boundary conditions are required to be imposed in general, giving in this way a complete physical meaning to these generators. In particular, the \textit{mass} is the value of the time-translation generator. The application of Regge-Teiteiboim method for asymptotically $n\ge 4$ anti-de Sitter spacetimes in presence of massive scalar fields was given in \cite{Henneaux:2006hk} 
and the massless case in \cite{Saenz:2012ga}.  For asymptotically flat spacetimes, the only contribution of a massless scalar field is the surface integral coming from its kinetic term. This contribution supplements the pure gravity contribution yielding the following expression for the variation of the mass $\cal M$ for the Fisher class \eqref{JNWsol}-\eqref{JNWsolphi},  
\begin{align} \label{varFishermass}
\delta {\cal M}=\left( \pm \frac{1}{2}  \sqrt{\frac{(n-2) (n-3) \left(1-\alpha ^2\right) \epsilon }{\kappa_n }} M \delta \phi_0 + \frac{(n-2) }{2\kappa_n} \delta(\alpha M )\right)\Omega_{n-2}. 
\end{align}
To obtain the mass from its variation \eqref{varFishermass} is required either i) a functional relation between $\phi_0$, $M$ and $\alpha$ or ii) to set $\delta \phi_0=0$. Hereafter, the Dirichlet boundary condition $\delta \phi_0=0$, i.e.,  {that where} the scalar field is fixed at infinity,  is adopted. Under this condition the mass is given by
\begin{align} \label{Fishermass}
{\cal M}= \frac{(n-2) }{2\kappa_n}\alpha M \Omega_{n-2}.
\end{align}
Note that although the expressions \eqref{MADM} and \eqref{Fishermass}  coincide, they correspond to different boundary conditions: the first one assumes  a vanishing scalar field at infinity ($\phi_0=0$), while the second requires only to fix the the scalar field at infinity ($\delta\phi_0=0$).

The spacetime (\ref{JNWsol}) reduces to  the Schwarzschild-Tangherlini metric for $\alpha^2=1$ by the Birkhoff's theorem. The metric (\ref{JNWsol}) for $\alpha=-1$ differs in disguise from the standard Schwarzschild-Tangherlini metric. This can be remedied by a coordinate transformation $S=(r^{n-3}-M)^{1/(n-3)}$, leading to 
\begin{align}
\D s^2=&-\biggl(1+\frac{M}{S^{n-3}}\biggl)\D { t}^2+ \biggl(1+\frac{M}{S^{n-3}}\biggl)^{-1}\D S^2+S^2\D \Omega_{n-2}^2.
\end{align}
This corresponds to the Schwarzschild-Tangherlini solution whose mass parameter has an opposite sign. 
Namely, the sign flip of $\alpha$ is compensated by that of the mass parameter. 

The same remark applies also to the $\alpha^2 \ne 1$ case. 
One sees immediately that under the simultaneous transformation
\begin{align}
\label{refrection}
M\to -M, \qquad \alpha\to -\alpha,
\end{align}
the metric (\ref{JNWiso}) is left unaltered and the scalar field varies as $\phi-\phi_0 \to -(\phi-\phi_0)$. 
Since the sign change of $\phi-\phi_0$ does not affect the the causal structure of spacetime, 
this freedom allows us to focus only on the $\alpha >0$ case. 
However, we shall not attempt to put this restriction for the easy comparison with the 
results in the literature. 

A distinguished case to be observed is the $\alpha=0$ case, for which the {mass} 
vanishes. The metric is ``ultra-static'' ($g_{tt}=-1$) but is not isometric to the Minkowski spacetime. 
This is a prominent example for which the positive mass theorem is not applied because of the 
violation of the null energy condition.
The other {${\cal M}=0$} case is achieved by $M=0$,  recovering the Minkowski spacetime. 
Hereafter, we assume $M\ne 0$ and $\alpha^2\ne 1$.

\subsubsection{Areal radius and proper volume}

The areal radius $S$ is given by 
\begin{align}
S(r)=rf(r)^{(1-\alpha)/[2(n-3)]}.
\end{align}
Hence, for $M>0$, $S$ becomes zero {for $\alpha<1$ and matches $r_{\rm s}$ if  $\alpha=1$} in the limit of $r\to r_{\rm s}$, while it diverges  for $\alpha>1$ as $r\to r_{\rm s}$.
For $M<0$, in the limit of $r\to 0$, $S$ becomes zero for $\alpha>-1$, is finite for {$\alpha=-1$}, while it diverges for $\alpha<-1$.

The proper volume of the domain from $r=r_0$ satisfying $0<f(r_0)<\infty$ to $r>r_0$ on a spacelike hypersurface with constant $t$ is given by 
\begin{align}
{\rm Vol}(r)=\Omega_{n-2}\left|{\int_{r_0}^r r^{n-2}f(r)^{[2-(n-1)\alpha]/[2(n-3)]}\D r}\right|.
\end{align}
For $M>0$, ${\rm Vol}(r)$ is divergent for $\alpha \ge 2(n-2)/(n-1)$ in the limit of $r\to r_{\rm s}$, while it is finite for $\alpha<2(n-2)/(n-1)$.
For $M<0$, ${\rm Vol}(r)$ becomes zero for $\alpha>  -2(n-2)/(n-1)$ in the limit of $r\to 0$, while it diverges for $\alpha\le -2(n-2)/(n-1)$.

\subsubsection{Radial non-timelike geodesics}

Let us consider an affinely-parametrized radial null geodesic $\gamma$ with its tangent vector $k^\mu(=\D x^\mu/\D \lambda)$, where $\lambda$ is an affine parameter for $\gamma$.
By the existence of a Killing vector $\xi^\mu=(\partial/\partial {t})^\mu$, 
$C=-k_\mu \xi^\mu(>0)$ deserves a constant of motion--corresponding to the energy of the 
null particle--along $\gamma$.
Combining this with $k_\mu k^\mu=0$, we obtain
\begin{align}
\label{JNWradialnull}
\frac{\D r}{\D \lambda}=\pm Cf(r)^{(n-4)(1-\alpha)/[2(n-3)]},
\end{align}
where $\pm$ sign represents the outgoing and ingoing geodesics.
This equation shows
\begin{align}
\int f(r)^{(n-4)(\alpha-1)/[2(n-3)]}\D r=\pm C \lambda.
\end{align}
If $\lambda$ diverges at some value of $r$, infinite amount of affine time 
elapses for a radial null geodesic to attain that surface. In short, it corresponds to 
{\it null infinity}.

Let us consider the $r=r_{\rm s}$ surface for $M>0$, around which $f(r)\simeq f'(r_{\rm s})(r-r_{\rm s})$ holds.
In $n=4$, one finds that $r=r_{\rm s}$ is not null infinity insensitive to the value of $\alpha$.
In contrast, $r=r_{\rm s}$ is null infinity for $\alpha \le -(n-2)/(n-4)$ in higher dimensions ($n\ge 5$).
We would like to emphasize that this null infinity 
($\alpha \le -(n-2)/(n-4)$ with $n\ge 5$) has a vanishing area. 
On top of this, the area diverging surface ($r=r_{\rm s}$ with $\alpha>1$) can 
be reached within a finite affine time for $\gamma$. 
The readers may feel that this is counter-intuitive, as these 
eccentric behaviors do not occur  e.g., in the Schwarzschild spacetime. 
This peculiar feature highlights the spacetime with a phantom scalar field. 

To see this more concretely, let us examine the expansion rate 
$\theta_-=k^\mu \nabla_\mu (S^{n-2})/S^{n-2}$ 
for the ingoing radial null geodesics (set $C=1$ and choose the minus sign 
in (\ref{JNWradialnull})): 
\begin{align}
\label{}
\theta_- =-(n-2)f^{(n-4)(1-\alpha)/[2(n-3)]} \left(\frac{1}{r}+\frac{1-\alpha}{2(n-3)}\frac{\partial_r f}{f}\right).
\end{align}
For the Minkowski spacetime $f(r)=1$, $\theta_-$ is negative-definite, implying that the light rays are focusing. 
In the $\alpha>1$ case, 
to the contrary, the second term in the bracket for the above equation 
changes sign, allowing $\theta_-$ to have a positive value. 
This means that the area is increasing even along the ``ingoing'' null geodesics. 
This property attributes to the violation of the null energy condition, 
which controls $k^\mu\nabla_\mu \theta_-$ through Raychaudhuri's equation.

Next we consider $r=0$ for $M<0$, around which $f(r)\simeq {\cal O}(r^{-(n-3)})$ holds.
For $n=4$, $r=0$ is not null infinity, independent of the value of $\alpha$.
In contrast, $r=0$ is null infinity for $\alpha \ge (n-2)/(n-4)$ for $n\ge 5$.

Let us check the spacelike distance as well.
Along a radial spacelike geodesic on a spacelike hypersurface with constant $t$, the proper distance from $r_0$ to $r$ is given by 
\begin{align}
s=\left|\int^r_{r_0} f(r)^{-(\alpha+n-4)/[2(n-3)]}\D r\right|,
\end{align}
If $s$ diverges at some value of $r$, one measures an infinite affine length 
from the ambient space to reach that surface, i.e., it corresponds to {\it spacelike infinity}.

For $M>0$, $r=r_{\rm s}$ is spacelike infinity for $\alpha\ge n-2$.
For $M<0$, $r=0$ is spacelike infinity for $\alpha\le -(n-2)$.

\subsubsection{Singularity}

The spacetime singularity is usually characterized by the divergence of 
spacetime curvature invariants. 
For the Fisher metric (\ref{JNWsol}), the Ricci scalar, for instance, is computed as 
\begin{align}
\label{}
R=(n-2)(n-3)(1-\alpha^2)\frac{M^2}{4r^{2(n-2)}} f(r)^{(\alpha-n+2)/(n-3)}.
\end{align}
Let us first concentrate on the  $r=r_{\rm s}$ surface for the $M>0$ case. 
A quick consequence of the above expression of $R$ is that $r=r_{\rm s}$ surface is singular when $\alpha <n-2$. 
The divergence behavior of the Kretschmann scalar is the same as that of $R$.

For $\alpha>n-2$, scalar quantities constructed out of curvature tensors 
tend to be zero as $r\to r_{\rm s}$. One might be therefore inclined to expect that 
$r= r_{\rm s}$ is a regular surface. 
Let us point out that this is not a completely regular surface but a
p.p curvature singularity, which is characterized by the divergence of 
curvature tensor in a basis parallelly propagated along some curve $\gamma'$ \cite{Hawking:1973uf}.  It should be noted that a scalar curvature singularity is always a p.p curvature singularity, but the converse is not true. 
To demonstrate the existence of a p.p curvature singularity, 
let us devote ourselves to the radial null geodesic $\gamma$
as before. As shown in appendix \ref{app:geom}, 
one can construct pseudo-orthonormal frame ($k^\mu, n^\mu, E_{\hat i}{}^\mu$) 
as (\ref{nEi}), which is parallelly propagated along the radial null geodesic 
 $k^\nu \nabla_\nu k^\mu=k^\nu \nabla_\nu n^\mu=k^\nu \nabla_\nu E_{\hat i}{}^\mu=0$.
 Setting $f_1(r)=f(r)^\alpha$, $f_2(r)=f(r)^{-(\alpha+n-4)/(n-3)}$ and 
 $S(r)=rf(r)^{(1-\alpha)/[2(n-3)]}$ in (\ref{metricf1f2S}),  
the Riemann tensor component  (\ref{Rili}) in this frame is computed as 
\begin{align}
\label{JNW:Rlini}
R_{\mu\nu\rho\sigma}k^\mu E_{\hat i}{}^\nu k^\rho E_{\hat j}{}^\sigma 
=-\frac{(\alpha^2-1)(n-3)M^2}{4r^{2(n-2)}}f(r)^{-[(n-4)\alpha+(n-2)]/(n-3)} 
\delta_{\hat i\hat j},
\end{align}
One verifies that 
this component diverges as $r\to r_{\rm s}$ for $\alpha>-(n-2)/(n-4)$. 
This range of $\alpha$ covers the parameter region in which curvature invariant scalars fail to diverge. 
We thus conclude that $r= r_{\rm s}$ is a singular surface in the entire  parameter region. 

Let us next explore the causal nature of this singularity. 
For our current purpose, it is sufficient to focus on the 
two-dimensional Lorentzian portion of the spacetime 
\begin{align}
\D s_2^2=&f(r)^{\alpha}\biggl(-\D { t}^2+ f(r)^{-\{(n-2)\alpha+n-4\}/(n-3)}\D r^2\biggl)
=f(r(r_*))^{\alpha}(-\D { t}^2+ \D r_*^2),
 \label{2-dimJNW}
\end{align}
where we have introduced a tortoise coordinate 
\begin{align}
r_*:=\int f(r)^{p}\D r,\qquad \biggl(p:=-\frac{(n-2)\alpha+n-4}{2(n-3)}\biggl).
\label{rho-JNW}
\end{align}
If some value of $r$ corresponds to an finite (infinite) value of $r_*$, it enjoys a timelike (null) structure 
in the Penrose diagram.
Expanding $f(r)\simeq f'(r_{\rm s})(r-r_{\rm s})$ around $r=r_{\rm s}$, one sees immediately that 
the singularity $r=r_{\rm s}$ is timelike for $\alpha< 1$, and null for $\alpha \ge 1$.

Our results for $M>0$ case is summarized in Table \ref{table:JNW1}.

\begin{table}[htb]
\begin{center}
\caption{\label{table:JNW1} Properties of the singularity $r=r_{\rm s}$ in the Fisher spacetime with $M>0$ and $\alpha^2\ne 1$. Note $-(n-2)/(n-4)\to -\infty$ for $n=4$.}
\vspace{0.2cm}
\begin{tabular}{|c|c|c|c|c|c|}
\hline
 & $\alpha\le -\frac{n-2}{n-4}$ & $-\frac{n-2}{n-4}<\alpha<1$ & $1<\alpha<\frac{2(n-2)}{n-1}$ & $\frac{2(n-2)}{n-1}\le\alpha<n-2$ & $ n-2\le \alpha$  \\ \hline\hline
Type &Scalar &Scalar & Scalar & Scalar  & p.p.  \\ \hline 
Signature & Timelike& Timelike & Null & Null & Null  \\ \hline 
Null infinity? & Yes & No & No& No & No  \\ \hline
Spacelike infinity? & No & No & No & No & Yes   \\ \hline
Area & 0 & 0 & $\infty$ & $\infty$ & $\infty$   \\ \hline
Volume & Finite & Finite & Finite & $\infty$ & $\infty$   \\ \hline
\begin{tabular}{c}
Penrose diagram \\
 in Fig. \ref{fig:PDJNW}
\end{tabular}
& (I)& (II) & (III) & (III) & (IV) \\ \hline
\hline
\end{tabular} 
\end{center}
\end{table} 

\bigskip
We next turn our attention to the $M<0$ case, for which  
 the domain of $r$ is $0<r<\infty$.
Around $r=0$, we have $f(r)\simeq {\cal O}(r^{-(n-3)})$, giving 
\begin{align}
\lim_{r\to 0}{R}\propto r^{-(\alpha+n-2)}. \label{JNW-Ricci-r0}
\end{align}
Thus, $r=0$ is a scalar curvature singularity in the case of $\alpha> -(n-2)$.
From (\ref{JNW:Rlini}), one sees that 
$R_{\mu\nu\rho\sigma}k^\mu E_{\hat i}{}^\nu k^\rho E_{\hat j}{}^\sigma \simeq r^{(n-4)\alpha-(n-2)}$ 
at $r=0$, leading to the conclusion that $r=0$ is a p.p curvature singularity for 
$\alpha{<} (n-2)/(n-4)$. It is therefore found that $r=0$ is singular, irrespective of the value of 
$\alpha$. 

For the signature of the singularity at $r=0$, 
Eq.~(\ref{rho-JNW}) shows that $r=0$ is timelike 
for $\alpha>-1$ and null for $\alpha \le -1$.

We encapsulate the properties of singularity at $r=0$ in Table \ref{table:JNW2}.
One verifies that the singularity structure of $r=r_{\rm s}$ and $r=0$ is 
symmetric under $\alpha \to -\alpha$, consistent with the symmetry (\ref{refrection}). 

\begin{table}[htb]
\begin{center}
\caption{\label{table:JNW2} Properties of the singularity $r=0$ in the Fisher spacetime with $M<0$ and $\alpha^2\ne 1$. Note $(n-2)/(n-4)\to +\infty$ for $n=4$.}
\vspace{0.2cm}
\scalebox{0.9}{
\begin{tabular}{|c|c|c|c|c|c|}
\hline
 & $\alpha\le -(n-2)$ & $-(n-2)<\alpha\le -\frac{2(n-2)}{n-1}$ & $-\frac{2(n-2)}{n-1}<\alpha<-1$ & $-1<\alpha<{\frac{n-2}{n-4}}$ & $\alpha\ge \frac{n-2}{n-4}$ \\ \hline\hline
Type & p.p. & Scalar  & Scalar  & Scalar  & Scalar \\ \hline 
Signature & Null & Null & Null & Timelike  & Timelike  \\ \hline  
Null infinity? & No & No & No & No  & Yes  \\ \hline
Spacelike infinity? & Yes & No  & No & No  & No  \\ \hline
Area & $\infty$ & $\infty$ & $\infty$ & $0$  & $0$  \\ \hline
Volume  & $\infty$ & $\infty$ & $0$ & $0$  & $0$  \\ \hline
\begin{tabular}{c}
Penrose diagram \\
 in Fig. \ref{fig:PDJNW}
\end{tabular}
& (IV)& (III) & (III) & (II)  & (I) \\ \hline
\hline
\end{tabular} 
}
\end{center}
\end{table} 

\subsubsection{Penrose diagrams}

From the properties of singularity structure and the behavior of null geodesics, 
one can depict the global structure of the spacetime. 
A convenient device suitable for the visualization of global causal structure 
is the Penrose diagram, which compactifies the spacetime while keeping 
light ray propagations conformally invariant. 

\begin{figure}[t]
\begin{center}
\includegraphics[width=10cm]{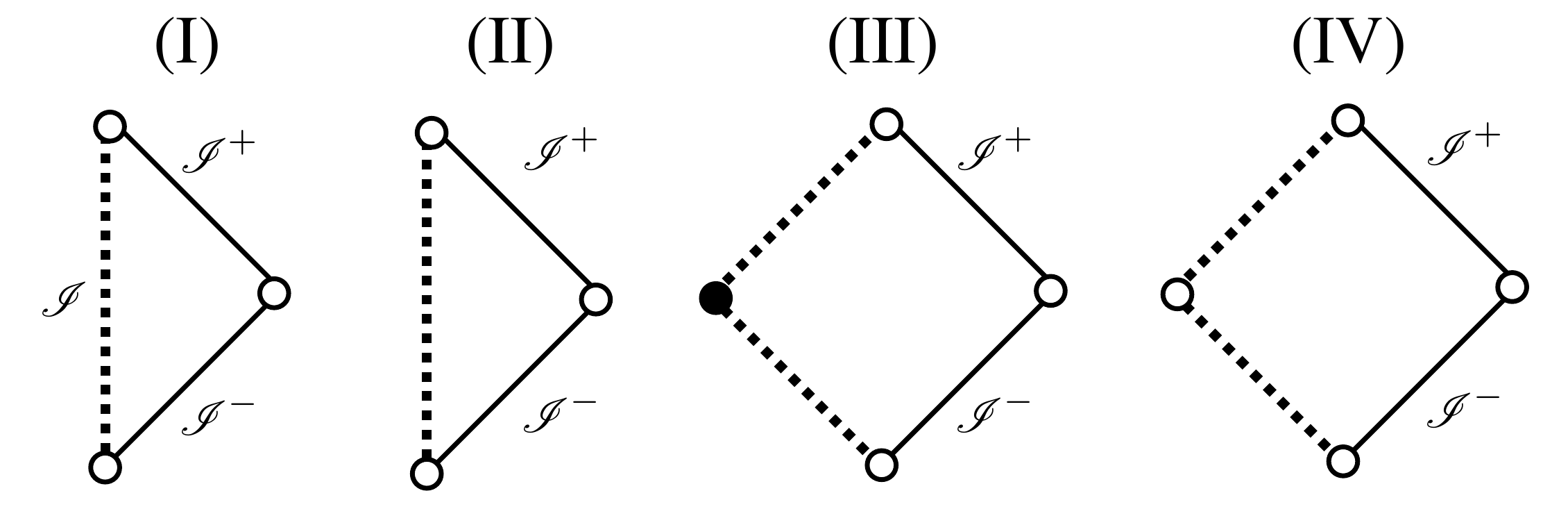}
\caption{Possible conformal diagrams for the spherical Fisher and Ellis-Gibbons solutions.
The dashed lines correspond to the (scalar or p.p) singularities. 
$\mas I^\pm $ represent the future/past null infinities, whereas the timelike boundary
$\mas I$ in (I) also corresponds to infinity. 
White and black circles stand  for spatial/timelike infinities, 
and for the singular bifurcation, respectively.}
\label{fig:PDJNW}
\end{center}
\end{figure}

The Penrose diagrams for the phantom Fisher solution are shown in Figure \ref{fig:PDJNW}. 
We obtained four different causal structures. The correspondence to the each diagram 
and the parameter region of $\alpha$ has been already specified in Table \ref{table:JNW1} and \ref{table:JNW2}.

\subsubsection{Positive mass and violation of the energy conditions}

Readers may wonder why the phantom Fisher solution admits a positive ADM mass, in spite of the 
violation of the energy conditions everywhere. To figure out this, it is instructive to see 
the Komar integral associated with the static Killing vector $\xi^\mu=(\partial/\partial t)^\mu$: 
\begin{align}
\label{Komar}
\frac 1{16\pi}\int _{\partial \Sigma} \nabla_\mu \xi_\nu \D S^{\mu \nu}
=\frac 1{8\pi} \int_\Sigma R_{\mu\nu}\xi^\mu \D \Sigma^\nu.
\end{align}
Here we have denoted $\D \Sigma_\mu =- u_\mu \D \Sigma$ and $\D S^{\mu\nu}=-2u^{[\mu}n^{\nu]}\D S$, 
where $u_\mu$ is a future pointing unit normal to the hypersurface $\Sigma=\{t={\rm const.}\}$ 
and $n_\mu$ is an outward pointing unit normal to the boundary $\partial \Sigma$ in $\Sigma$. 
Since we are considering the static massless scalar system (\ref{em-kg}), 
we have the Ricci staticity $R_{\mu\nu}\xi^\mu =0$, rendering the right hand side of (\ref{Komar})
to vanish. It therefore follows that the contribution from infinity $\partial\Sigma_\infty$ (giving rise to the ADM mass) must offset the contribution coming from interior boundary 
$\partial\Sigma_{\rm int}$ ($r=r_{\rm s}$ surface in the present case) of $\Sigma$. 
It therefore follows that the presence of two boundaries--and thus the presence of  image charge at 
$\partial\Sigma_{\rm int}$--is a primary reason for the existence of the positive ADM mass configuration.

\subsection{Ellis-Gibbons class}

In the spherically symmetric case ($\epsilon =-1$ and $k=1$ with $\D \Sigma_{k=1,n-2}^2$ being the 
metric of round sphere), the Ellis-Gibbons solution reads
\begin{align}
\label{Gibbons:spherical}
\D s^2=&-e^{-M/r^{n-3}}\D { t}^2+e^{M/[(n-3)r^{n-3}]}\biggl(\D r^2+r^2\D \Omega_{n-2}^2\biggl),\\
\phi=&\phi_0\pm \sqrt{-\epsilon\frac{n-2}{4(n-3)\kappa_n}}\frac{M}{r^{n-3}}.
\end{align}
The solution is manifestly asymptotically flat for $r\to \infty$. Following the guidelines introduced in section \ref{section-mass}, the mass of the Ellis-Gibbons class is determined as 
\begin{align}
\label{}
{\cal M}=\frac{(n-2) \Omega_{n-2} }{2\kappa_n}M.
\end{align}

Our next concern is the coordinate singularity at the origin $r=0$, which 
we turn to discuss. The entire argument is parallel with that for the Fisher solution. 
In the subsequent analysis, 
we will utilize the following integral 
\begin{align}
\label{incompGammaformula}
\int_{r}^{\infty}  r^{p} e^{-Mq/r^{n-3}} \D r=-\frac{(Mq)^{(1+p)/(n-3)}}{n-3}\Gamma \left(
-\frac{1+p}{n-3}, \frac{M q}{r^{n-3}} 
\right),
\end{align}
where $p$, $q$ are constants, and 
$\Gamma(a,z)$ is an incomplete gamma function defined by
$\Gamma(a,z)=\int^{\infty}_z t^{a-1}e^{-t}\D t$. 
The asymptotic expansion as $|z|\to \infty$ is given by  
\begin{align}
\label{incGamma:asy}
\Gamma(a,z)\simeq e^{-z}z^{a-1}, \quad (z\gg 1), \qquad 
\Gamma(a,z)\simeq e^{-z+2i(a-1)\pi}(1/z)^{1-a}, \quad (z\ll -1), 
\end{align}
Namely the integral (\ref{incompGammaformula}) is divergent as $r\to 0$ for $Mq<0$, 
while it converges to a finite value as $r\to 0$ for $Mq\ge 0$.

\subsubsection{Areal radius and proper volume}

The areal radius is given by $S:=re^{M/[2(n-3)r^{n-3}]}$. 
For $M>0$, the areal radius diverges as $r\to 0$, 
while it becomes zero as $r\to 0$ for $M<0$. 

The proper volume of the domain $r\in [r_0, r]$ on a static timeslice is given by
\begin{align}
\label{}
{\rm Vol}(r)&= \Omega_{n-2} \left|\int_{r_0}^r r^{n-2}e^{(n-1)M/[2(n-3)r^{n-3}]} \D r \right|.
\end{align}
This integral can be evaluated by setting $p=n-2$ and $q=-(n-1)/[2(n-3)]$ in (\ref{incompGammaformula}). 
One finds the proper volume is divergent for $M>0$ in the $r\to 0$ limit, 
while it tends to be zero for $M<0$ as $r\to 0$.

\subsubsection{Radial non-timelike geodesics}

The tangent vector for the radial null geodesics in the Ellis-Gibbons solution 
(\ref{Gibbons:spherical}) is given by 
$k^\mu = Ce^{M/r^{n-3}}(\partial/\partial t)^\mu\pm C e^{(n-4)M/[2(n-3)r^{n-3}]}(\partial /\partial r)^\mu$, 
where $C$ is a geodesic constant. We are therefore led to 
\begin{align}
\label{}
\pm C \lambda &={\int}  e^{-(n-4)M/[2(n-3)r^{n-3}]} \D r.
\end{align}
Substituting $p=0$ and $q=(n-4)/[2(n-3)]$ in (\ref{incompGammaformula}), 
one finds that $r=0$ is not null infinity in $n=4$ for either sign of $M$, 
and in $n\ge 5$ for $M>0$.  
In $n\ge 5$ with $M<0$, $r=0$ corresponds to the null infinity.

The proper distance along a constant timeslice is computed to
\begin{align}
\label{}
s&=\left|{\int} e^{M/[2(n-3)r^{n-3}]} \D r \right|.
\end{align}
Setting $p=0$ and $q=-1/[2(n-3)]$ in (\ref{incompGammaformula}), 
this blows up for $M>0$ as $r\to 0$, i.e., 
the $r=0$ surface corresponds to the spacelike infinity for $M>0$. 
For $M<0$, the proper distance to the $r=0$ surface is finite. 

\subsubsection{Singularity}

Let us next explore the singular behavior at $r=0$. 
The Ricci scalar is 
\begin{align}
\label{}
R=-\frac{(n-2)(n-3)}{4 r^{2(n-2)}}M^2 \exp \left(-\frac{M}{(n-3) r^{n-3}}\right).
\end{align}
In the $M<0$ case, $r=0$ is an obvious scalar curvature singularity. 
For $M>0$, $R$ does not blow up at the $r=0$ surface. 
All other curvature invariants seem to be finite there. 
However, this is a p.p curvature singularity  since
(\ref{Rili}) gives rise to 
\begin{align}
\label{}
R_{\mu\nu\rho\sigma}k^\mu E_{\hat i}{}^\nu k^\rho E_{\hat j}{}^\sigma 
=-\frac{(n-3) M^2}{4r^{2(n-2)}} \exp\left(\frac{(n-4)M}{(n-3)r^{n-3}}\right)\delta_{\hat i\hat j},
\end{align}
which tends to diverge as $r\to 0$ in arbitrary $n\ge 5$ for $M>0$, and for $n=4$ with any $M\ne 0$. 
It follows that $r=0$ is singular in any parameter region.

The two-dimensional portion of the metric is 
\begin{align}
\label{}
\D s_2^2=-e^{-M /r^{n-3}}\left(\D t^2-e^{(n-2)M/[(n-3)r^{n-3}]}\D r^2\right)=-e^{-M /r(r_*)^{n-3}}(\D t^2-\D r_*^2), 
\end{align}
where  the tortoise coordinate $r_*$ reads
\begin{align}
\label{}
r_*=\int e^{(n-2)M/[2(n-3)r^{n-3}]}\D r,
\end{align}
corresponding to $p=0$ and $q=-(n-2)/[2(n-3)]$ in (\ref{incompGammaformula}). 
It thus follows that $r=0$ surface is null for $M>0$, while 
it is timelike for $M<0$. 

\subsubsection{Penrose diagrams}

Properties of the coordinate singularity at $r=0$ are summarized in Table \ref{table:Gibbons}. 
The Penrose diagram can be deduced by bringing the issues obtained above together, 
as shown in the last row in Table \ref{table:Gibbons}.   

The areal radius $S(r)=r e^{M/[2(n-3)r^{n-3}]}$ admits a minimum at $r_{\rm th}^{n-3}=M/2$ for $M>0$, which 
have led some authors to conclude that the spherical Ellis-Gibbons solution is a 
regular traversable wormhole with a throat at $r=r_{\rm th}$~\cite{Boonserm:2018orb}.\footnote{The assertion in \cite{Boonserm:2018orb}
does not directly counter to the uniqueness theorems of wormholes, 
since the Ellis-Gibbons solution does not uphold the prerequisites of the proofs in \cite{Yazadjiev:2017twg,Rogatko:2018smj},  which have assumed two asymptotically flat regions. 
The spherical Ellis-Gibbons solution is not asymptotically flat around $r=0$.} 
However, our analysis has explicitly demonstrated that  the Ellis-Gibbons metric describes an asymptotically flat singular solution in any parameter region.

\begin{table}[t]
\begin{center}
\caption{\label{table:Gibbons} Properties of the singularity $r=0$ in the 
spherical Ellis-Gibbons solution.}
\vspace{0.2cm}
\begin{tabular}{|c|c|c|c|}
\hline
 & $M>0$ & \multicolumn{2}{|c|}{$M<0$}  \\ \hline\hline
 Dimensions & {$n\ge 4$} & $n=4$ & $n\ge 5$ \\ \hline 
Type &p.p. &Scalar  & Scalar   \\ \hline 
Signature & Null & Timelike  & Timelike \\ \hline 
{Null infinity?} & No & No &  Yes  \\ \hline
{Spacelike infinity?} & Yes & No & No  \\ \hline
Areal & $\infty$ &0 & 0   \\ \hline
Volume & $\infty$ & 0& 0   \\ \hline
\begin{tabular}{c}
Penrose diagram \\
 in Fig. \ref{fig:PDJNW}
\end{tabular}
& (IV)& (II) & (I) \\ \hline
\hline
\end{tabular} 
\end{center}
\end{table} 

\subsection{Ellis-Bronnikov class} 
\label{sec:EB}

The spherical Ellis-Bronnikov class of solutions is given by 
\begin{align}
\label{Ellis-Bronnikov:spherical}
\D s^2&=-e ^{-2\beta U(r)}\D  t^2+e^{2\beta U(r)/(n-3)}V(r)^{1/(n-3)}
\left(\frac{\D r^2}{V(r)}+r^2 \D\Omega_{n-2}^2 \right), \\
\phi&=\phi_0\pm\sqrt{-\epsilon\frac{(n-2)(1+\beta^2)}{\kappa_n(n-3)}} U(r),
\\
U(r):&= \arctan \left(\frac{M}{2r^{n-3}}\right), \qquad 
V(r):= 1+\frac{M^2}{4r^{2(n-3)}}.
\end{align}
In four dimensions, this solution was first derived by Ellis \cite{Ellis1973} and 
Bronnikov \cite{Bronnikov1973}.
Defining the isotropic coordinate $\rho $ by
\begin{align}
\label{}
r=\rho \left(1-\frac{M^2}{16\rho^{2(n-3)}}\right)^{1/(n-3)},
\end{align}
the solution can be transformed into
\begin{align}
\label{Ellis-Bronnikov:iso}
\D s^2&=- e^{-2\beta \hat U(\rho)}\D t^2+ e^{2\beta \hat U(\rho)/(n-3)}\left(
1+\frac{M^2}{16\rho^{2(n-3)}} 
\right)^{2/(n-3)}(\D \rho^2+\rho^2 \D \Omega_{n-2}^2), \\
\phi&=\phi_0\pm\sqrt{-\epsilon\frac{(n-2)(1+\beta^2)}{\kappa_n(n-3)}} \hat U(\rho),
\end{align}
where $\hat U(\rho)=U(r(\rho))$. {Using the Dirichlet boundary condition $\delta\phi_0=0$ discussed in section \ref{section-mass}, the mass of  this class is given by }
\begin{align}
\label{EB:ADM}
{\cal M}=\frac{(n-2)\Omega_{n-2}}{2\kappa_n}\beta M.
\end{align}

It is important to recognize that the spacetime admits an inversion symmetry. 
In terms of 
\begin{align}
\label{}
\rho =\frac{(M^2/4^2)^{1/(n-3)}}{\hat \rho},
\end{align}
the metric can be cast into 
\begin{align}
\label{}
\D s^2&=- e^{2\beta \hat U(\hat \rho)}\D t^2+ e^{-2\beta \hat U(\hat \rho)/(n-3)}\left(
1+\frac{M^2}{16\hat \rho^{2(n-3)}} 
\right)^{2/(n-3)}(\D \hat \rho^2+\hat \rho^2 \D \Omega_{n-2}^2), \\
\phi&=\phi_0\mp \sqrt{-\epsilon\frac{(n-2)(1+\beta^2)}{\kappa_n(n-3)}} \hat U(\hat \rho).
\end{align}
This amounts to 
\begin{align}
\label{EG:signflip}
M \to - M, \qquad \beta \to -\beta.
\end{align}
Therefore, physical properties of the solution are unchanged under this sign flip. 

Analogous to the $\alpha=0$ Fisher solution, 
the $\beta=0$ case is noteworthy here, since it gives a nontrivial spacetime configuration 
even for the vanishing {mass}. In this case, the metric is ultra-static and the static observer with velocity vector $\partial/\partial t$ encounters no tidal force of the spacetime. This geometry has been fully studied in the literature, since it corresponds to the zero mass wormhole, see e.g. \cite{Bhattacharya:2010zzb,Abe:2010ap,
Tsukamoto:2016qro,Cremona:2018wkj,Roy:2019yrr} and references adduced therein. 
The other vanishing {mass} state $M=0$ corresponds to the Minkowski metric, 
which we shall exclude from our analysis. 

The solution (\ref{Ellis-Bronnikov:spherical}) admits a coordinate singularity at $r=0$, 
which we are going to study. 
In the following analysis, we need to evaluate the integral 
$\int_{r_0}^r r^{m}e^{p\beta U(r)}V(r)^{q} \D r$ around $r=0$. 
Approximating 
$\arctan[M/(2r^{n-3})]\simeq \pi {\rm sgn}(M)/2$ and 
$V(r)\simeq M^2/(4r^{2(n-3)})$ around $r=0$, 
we have 
\begin{align}
{\lim_{r\to 0}\int_{r_0}^r} r^{m}e^{p\beta U(r)}V(r)^{q} \D r \propto 
{\lim_{r\to 0}\int_{r_0}^r} r^{m-2(n-3)q} \D r.
\end{align}
Thus, this integral is divergent at $r=0$ for
\begin{align}
\label{Ellis:criteria}
m+1-2(n-3)q\le 0.
\end{align}

\subsubsection{Areal radius and proper volume}

The areal radius of the metric (\ref{Ellis-Bronnikov:spherical}) is given by 
$S(r)= re^{\beta U(r)/(n-3)}V(r)^{1/[2(n-3)]}$. For $r\to 0+$, 
the areal radius converges to the finite value 
$S(r)\to S_{\pm}=(4M^2)^{2/(n-3)}\exp(\pm \beta\pi/[2(n-3)])$ for $M \gtrless 0$.

The proper volume of the domain $0<r'<r$ on a spacelike hypersurface with constant $t$  is 
\begin{align}
\label{}
{\rm Vol}(r) =\Omega_{n-2} \left|\int  r^{n-2} e^{(n-1)\beta U/(n-3)} V(r)^{{1/(n-3)}} \D r\right|.
\end{align}
Using the criterion (\ref{Ellis:criteria}), the proper volume is kept finite as $r\to 0+$.

\subsubsection{Radial non-timelike geodesics}

The tangent vector for the radial null geodesics  of the 
Ellis-Bronnikov metric (\ref{Ellis-Bronnikov:spherical})
is given by
\begin{align}
\label{}
k^\mu=C e^{2\beta U} \left(\frac{\partial}{\partial t}\right)^\mu\pm C e^{(n-4)\beta U/(n-3)}V^{(n-4)/[2(n-3)]}\left(\frac{\partial}{\partial r}\right)^\mu,  
\end{align} 
where $C$ is a constant of motion along the geodesics. 
Hence the affine parameter reads 
\begin{align}
\label{}
\pm C \lambda ={\int_{r_0}^r} e^{-(n-4)\beta U/(n-3)}V^{-(n-4)/[2(n-3)]} \D r.
\end{align}
The criterion (\ref{Ellis:criteria}) implies that the affine parameter fails to diverge
as $r\to 0$, i.e., the $r=0$ surface is not null infinity. 

The affine distance along the radial spacelike geodesic on the constant $t $ is 
\begin{align}
\label{}
s= \left|\int_{r_0}^r e^{\beta U/(n-3)} V^{-(n-4)/[2(n-3)]} \D r \right|.
\end{align}
From (\ref{Ellis:criteria}), one concludes that the affine distance to $r=0$
along static timeslice is finite, so that the $r=0$ with $t={\rm const.}$ is not 
a spacelike infinity.

\subsubsection{Singularity}

The scalar curvature of the solution (\ref{Ellis-Bronnikov:spherical})  is given by
\begin{align}
\label{}
R=& -\frac {M^2}{4r^{2(n-2)}}(n-2)(n-3)(1+\beta^2)e^{-2\beta U/(n-3)}V^{-(n-2)/(n-3)}.
\end{align}
Since $U(r\to 0+)={\rm sgn}(M)\pi/ 2$, 
the Ricci scalar is finite as $r\to 0$. 
Similarly, all other scalar curvature invariants also remain finite at $r=0$. 

Taking the pseudo-orthonormal frame as in appendix \ref{app:geom}, 
let us look into the following Riemann tensor components 
\begin{align}
\label{}
R_{\mu\nu\rho\sigma}k^\mu E_{\hat i}{}^\nu k^\rho E_{\hat j}{}^\sigma 
=&- \frac{(n-3)(1+\beta^2)M^2}{4r^{2(n-2)}} e^{2(n-4)\beta U/(n-3)} V^{-(n-2)/(n-3)}, \\
R_{\mu\nu\rho\sigma}k^\mu E_{\hat i}{}^\nu n^\rho E_{\hat j}{}^\sigma 
=&\frac{(n-3)M}{8r^{2(n-2)}}[4\beta r^{n-3}+M(1-\beta^2)]e^{-2\beta U/(n-3)} V^{-(n-2)/(n-3)}.
\end{align}
Other components also do not show the blowing up behavior at $r=0$. 
We thus conclude that $r=0$ is merely a coordinate singularity.

\subsubsection{Throat structure and Penrose diagram}

Since the Ellis-Bronnikov metric (\ref{Ellis-Bronnikov:spherical}) is free of any singularities, 
it would deserve a wormhole solution, which we would like to elucidate in this subsection. 
To this end, let us first try to find the locus of the throat. 
There exist several distinct notions and definitions of a throat in the literature. 
Restricting to the static and spherically symmetric case, a widely accepted definition is the 
$(n-2)$-dimensional surface whose areal radius satisfies the flare-out condition \cite{Morris:1988cz,Hochberg:1998ha,Kim:2013tsa}. 

Since the spherically symmetric spacetimes do not generate gravitational waves, 
one is able to localize the gravitational energy. 
A useful quantity of this sort in spherical symmetry is the areal radius
$S= re^{\beta U(r)/(n-3)}V(r)^{1/[2(n-3)]}$ and 
the Misner-Sharp mass \cite{Misner:1964je,Hayward:1994bu} given by
\begin{align}
\label{}
M_{\rm MS}:&= \frac{(n-2)\Omega_{n-2}}{2\kappa_n}S^{n-3}\Bigl(1-(D S)^2 \Bigr)\notag \\
&= \frac{(n-2)\Omega_{n-2}}{2\kappa_n}e^{\beta U(r)} V(r)^{-1/2} \left(
M\beta -\frac{M^2(\beta^2-1)}{4r^{n-3}}
\right).
\end{align}
As $r\to \infty$, one sees that $M_{\rm MS}$ converges to the mass given in  (\ref{EB:ADM}).
In the context of wormhole, the Misner-Sharp mass plays the role of ``shape function'' 
\cite{Morris:1988cz} which controls the shape of the wormhole in the embedding space. 
The locus of the wormhole throat $r=r_{\rm th}$ should satisfy 
$(D S)^2=0$ as well as the ``flare-out condition'' \cite{Morris:1988cz,Hochberg:1998ha,Kim:2013tsa}
\begin{align}
\label{flareout}
\frac{\D }{\D S} \left(\frac{M_{\rm MS}}{S^{n-3}}\right)<0.
\end{align}
The geometric meaning of this condition is recognizable as follows. 
Let us write the metric of $(n-2)$-sphere as 
$\D \Omega_{n-2}^2=\D \theta^2+\sin^2\theta \D \phi^2+\cos^2\theta\D \Omega_{n-4}^2$ 
and consider the $\theta=\pi/2$ surface. Then we embed the 
two-dimensional surface $g_{SS}\D S^2+S^2\D \phi^2$
 into the flat Euclid space $\D S^2+\D Z^2+S^2 \D \phi^2$. 
Then, the flare-out condition (\ref{flareout}) requires that the throat is the neck of the curve $Z=Z(S)$.
We caution the reader to distinguish this condition from the simple minimal radius surface. 
To figure out this, 
let us work in the coordinate 
$\D s^2=-f_1(r)\D t^2+f_2(r)\D r^2+S^2(r) \D \Omega_{n-2}^2$, 
and suppose that $\partial_r (f_2^{-1})$ is finite at the throat $\partial_rS=0$.  
Then the simple flare-out condition (\ref{flareout}) reduces to 
$\partial_r^2 S>0$ at the throat, which means that $S_{\rm th}$ is nothing but a minimal radius. 
However, the finiteness of $\partial_r (f_2^{-1})$ at the throat is not always true, as we will see in the following.

Let us  first study the $\beta =0$ case, for which observers rest at constant $r$ follow geodesics.   
The areal radius $S= rV(r)^{1/[2(n-3)]}$ admits a critical point $\D S/\D r=0$ at 
$r=0$, at which $S_{{\rm th}, \beta=0}^{n-3}=|M|/2$. In this case 
the flare-out condition $\D (M_{\rm MS}/S^{n-3})/\D S=-(n-2)(n-3)\Omega_{n-2}/(\kappa_nS_{{\rm th}, \beta=0})<0$ is satisfied, so that $S_{{\rm th}, \beta=0}$ is a throat. Note that this does not coincide with 
the convexity condition of $S$, since the second derivative is computed to yield 
$\partial_r^2S=(2n-7)S_{{\rm th}, \beta=0}^{-(2n-7)}r^{2(n-4)}|_{r=0}$. 
Thus, the minimal condition $\partial_r^2 S>0$ is satisfied only for $n=4$, whereas 
the $n\ge 5$ is marginal  $\partial_r^2 S=0$. 
In any case, one can continue the spacetime across $r=r_{\rm th}$ 
to $r=0+$. As we have seen, $r=0$ is a regular surface, since it does not correspond to 
neither a scalar curvature singularity nor a p.p curvature singularity. 
Since the metric is invariant under $r\to -r$ for $\beta =0$,  
one can continue the spacetime across the $r<0$ region, which will be discussed below.

The $\beta \ne 0$ case asks for a more involved analysis. Thanks to the 
property (\ref{EG:signflip}), we can confine to the $M>0$ case hereafter. 
For $r>0$, the derivative of the areal radius $S= re^{\beta U(r)/(n-3)}V(r)^{1/[2(n-3)]}$
is $\partial_r S\propto V(r)^{-(2n-7)/[2(n-3)]}(1-M\beta r^{3-n}/2)$. 
When the  {mass} is positive  ($\beta>0$), 
there exists a critical point $\D S/\D r=0$ at 
\begin{align}
\label{throatradius}
{r=r_{\rm th}:=}\left(\frac 12 M \beta\right)^{1/(n-3)}.
\end{align}
At this point, we have 
$S_{{\rm th}, \beta>0}^{n-3}=\frac 12 M\beta \sqrt{1+\beta^{-2}} e^{\beta \arctan(1/\beta)} $ 
and the flare-out condition reduces to 
\begin{align}
\label{}
\left.\frac{\D}{\D S}\left(\frac{M_{\rm MS}}{S^{n-3}}\right)\right|_{r=r_{\rm th}} =-\frac{(n-2)(n-3)\Omega_{n-2} }{\kappa_n S_{{\rm th}, \beta >0}}<0, 
\end{align}
implying $S_{{\rm th}, \beta>0}$ is a throat. 
When the {mass} is negative ($\beta <0$), there appear no critical points $\partial_r S=0$ for $r\ge 0$ in $n=4$, while $r=0$ deserves a throat in $n\ge 5$ at $S_{{\rm th}, \beta=0}$.

Our next concern is the spacetime extension across $r=0$. 
For this analysis, the present coordinate system (\ref{Ellis-Bronnikov:spherical}) is not of use 
by the following two reasons. First, 
$U(r)=\arctan(M/(2r^{n-3}))$ is not continuous at $r=0$. 
As commented in footnote \ref{FNarctan}, 
this can be remedied if one replaces the function $U(r)=\arctan(M/(2r^{n-3}))$ appearing in the metric and the scalar field by $ \tilde U(r):=\pi/2 -\arctan(2r^{n-3}/M)$ (recall $M>0$). 
Still, the metric is ill-behaved at $r=0$ in higher dimensions, since 
we have $g^{rr}=\mathcal O(1/r^{2(n-4)})$ around $r=0$.  One can circumvent this second drawback by means of the new coordinate $x:=r^{n-3}$. To sum up, the metric and the scalar field for the 
Ellis-Bronnikov solution (\ref{Ellis-Bronnikov:spherical}) can be recast into 
\begin{subequations}
\label{EBx}
\begin{align}
\D s^2=&- e^{-2\beta U_x(x)}\D t^2+e^{2\beta U_x(x)/(n-3)} \left(
\frac{\D x^2}{(n-3)^2 V_x(x)^{(n-4)/(n-3)}}+V_x(x)^{1/(n-3)} \D \Omega_{n-2}^2 
\right),  \\
\phi=&\phi_0\pm\sqrt{-\epsilon\frac{(n-2)(1+\beta^2)}{\kappa_n(n-3)}} U_x(x),
\end{align}
\end{subequations}
where 
\begin{align}
\label{}
U_x(x):= \frac {\pi}2 -\arctan\left(\frac{2x}{M}\right), \qquad 
V_x(x):=x^2+\frac{M^2}{4}.
\end{align}
Since the every component of the metric and its inverse, and the scalar field given in (\ref{EBx}) are completely smooth 
at $x=0$, one can extend the spacetime into the $x<0$ region. 
Setting $\ti x=-e^{\pi \beta} x$ and expanding the metric around $\ti x\to \infty$, 
we have $U_x\simeq \pi -e^{\pi \beta}M/(2\ti x)$, leading to
\begin{align}
\label{}
\D s^2_{x\to -\infty} \simeq -e^{-2\pi\beta} \left(
1+\frac{M\beta e^{\pi \beta}}{\ti x}
\right)\D t^2+\left(
1-\frac{M\beta e^{\pi \beta}}{(n-3)\ti x}
\right)\left(\frac{\D \ti x^2}{(n-3)^2 \ti x^{2(n-4)/(n-3)}}+\ti x^{2/(n-3)} \D \Omega_{n-2}^2\right),
\end{align}
Defining $\ti t:=e^{-\pi\beta}t$ and $\ti r^{n-3}=\ti x$, this is the standard fall-off behavior of the metric 
in the asymptotically flat spacetime. 
This means that the spacetime attached to 
$x<0$ region is also asymptotically flat with the {mass} given by
\begin{align}
\label{maxex:ADMmass}
{\cal M}_{x<0} 
=-\frac{(n-2)\Omega_{n-2}e^{\beta \pi}}{2\kappa_n}\beta M.
\end{align}
In $n=4$, this gives 
${{\cal M}_{x<0}}=-4\pi \beta M e^{\beta \pi}/\kappa_4$, recovering Ellis' results \cite{Ellis1973}.

As far as the global structures are concerned, 
there are some notable differences  from the $\beta =0$ case. 
First of all, the gluing surface $r=0$ does not always correspond to the throat or minimal radius. 
According to our conventional intuition, one may visualize the situation in which two disjoint universes are connected at the throat. 
This orthodox picture seems unpalatable. 
From the viewpoint of maximal extension of the spacetime metric, 
the natural boundary of the one-sided universe is the coordinate singularity at $r=0$,
rather than the critical point of the areal radius. 
Secondly, the maximal extension is not right-left symmetric across $x=0$. 
Specifically, the sign of {mass}  for $x>0$ and $x<0$ regions is always opposite for $\beta \ne 0$. 
These eccentric facets have not been underscored so seriously in the past literature.

\begin{figure}[t]
\begin{center}
\includegraphics[width=5cm]{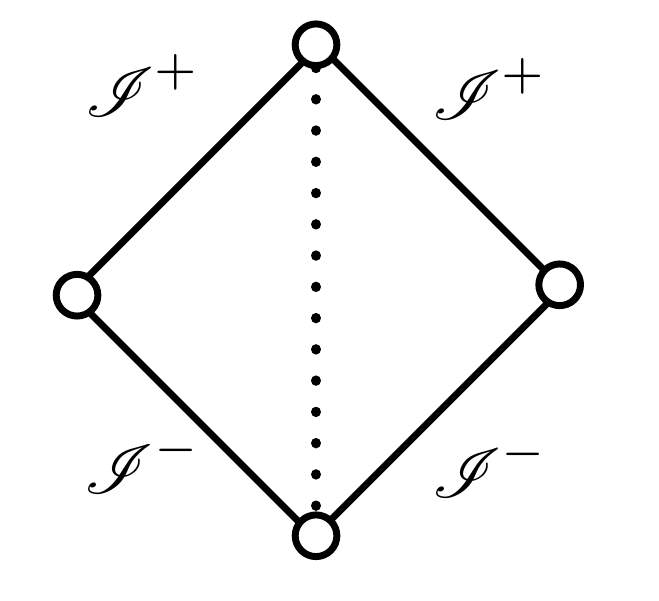}
\caption{Conformal diagram for the spherical Ellis-Bronnikov solution.
Solid lines denote null infinities $\mas I^\pm$ and dotted line corresponds to the $x=r^{n-3}=0$ surface, which 
does not always coincide with the throat for $\beta \ne 0$. In any parameter regions, 
this wormhole is traversable.}
\label{fig:PDEG}
\end{center}
\end{figure}

Let us display the global causal structure of the spherical Ellis-Bronnikov solution
in Figure \ref{fig:PDEG}. As we argued in this subsection, throat structures 
are different depending on the positivity of {mass}. 
In spite of this, the extensions through $x=0$ can be performed without any difficulty, and 
the maximal extensions of conformal diagram are all identical.

\section{Conformally coupled scalar field} 
\label{conformal}

In section \ref{sec:classification}, we have systematically classified all 
possible static solutions of warped form in Einstein-massless scalar system
for both signs of the kinetic term. Here, we proceed with the same classification in the case of  a conformally coupled massless scalar field. We will face this task by means of two methods. It is well known that there is a transformation which relates the solutions in the Einstein and Jordan frames, namely, the minimally and conformally coupled scalar fields.\footnote{The transformation was first found by J. Bekenstein \cite{Bekenstein:1974sf} for $n=4$ and $\varepsilon=1$. The generalization in \textit{arbitrary dimensions} can be found as a particular case of the general coupling $F(R,\phi)$ presented by K-i. Maeda \cite{Maeda:1988ab}. The specific case of a conventional conformally coupled scalar field was addressed by Xanthopoulos and Dialynas \cite{Xanthopoulos:1992fm}.} With this transformation along with the classification presented in section \ref{sec:classification}, we are able to obtain the catalogue of all possible static solutions of the same warped form in a system with a conformally coupled scalar field. We revisit two variants for obtaining this transformation, which are proved to be equivalent.  Indeed, the transformation is necessarily separated in two cases providing two branches of solutions. Of course, the second method is to solve the field equations in the Jordan frame. The latter approach is usually avoided since the rather involved field equations are required. However, Klim\v{c}\'{\i}k\cite{Klimcik:1993cia} was able to solve them in the case of a spherically symmetric conventional scalar field in arbitrary dimensions. Below, we introduce a generalization of Klim\v{c}\'{\i}k result, by obtaining the general solution assuming an arbitrary Einstein space $K^{n-2}$. The key point in this derivation is the choice of a suitable Ansatz for the metric.

Let us start our discussion for the non-minimally coupled scalar field
described by the action
\begin{align}
\label{actionconf}
S=\int \D ^n x \sqrt{-g}\left(\frac 1{2\kappa_n}
R -\frac {\varepsilon}2 (\nabla \Phi)^2 -\frac 12 \varepsilon\zeta R 
\Phi^2 \right),
\end{align}
where $\varepsilon=\pm 1$ and $\zeta $ is a dimensionless constant. 
The field equations derived from the above action are given by
\begin{align}
\label{CCscalar}
 \square \Phi -\zeta R\Phi=0,
\end{align}
and 
\begin{align}
\label{Eineq:co}
(1-\varepsilon \zeta \kappa_n \Phi^2)\left(R_{\mu\nu}
-\frac 12 R g_{\mu\nu}\right)=&\varepsilon \kappa_n \Biggl(2 \zeta g_{\mu\nu}\Phi  \square \Phi -2\zeta \Phi  \nabla_\mu \nabla_\nu \Phi
\notag \\ & +(1-2\zeta)(\nabla_\mu \Phi)(\nabla_\nu \Phi)
+\left(2\zeta -\frac 12 \right)g_{\mu\nu}(\nabla\Phi)^2 \Biggr).
\end{align}
{Note that for a generic \textit{constant} scalar field the field equations reduce to $R_{\mu\nu}=0$. For the special constant values $\varepsilon \zeta \kappa_n \Phi^2=1$, which are possible only for the conventional scalar field, the equations become a single one: $R=0$. In what follows, we discard solutions with a constant scalar field.}

Here we postulate that the scalar field equation (\ref{CCscalar}) is conformally invariant. 
This restricts $\zeta $ to be
\begin{align}
\label{}
\zeta =\zeta_c :=\frac{n-2}{4(n-1)}.
\end{align}
At this coupling constant, the trace of (\ref{Eineq:co}) together with (\ref{CCscalar}) 
entails a vanishing of the scalar curvature $R =0$ in the conformal frame. 
The Einstein's field equations simplify to
\begin{align}
\label{EOMcc}
E_{\mu\nu}:=R_{\mu\nu}-\varepsilon \kappa_n S_{\mu\nu}=0, \qquad \square \Phi=0,
\end{align}
where 
\begin{align}
\label{}
S_{\mu\nu} :=& \frac{n\nabla_\mu \Phi\nabla_\nu \Phi-g_{\mu\nu}(\nabla\Phi)^2
-(n-2)\Phi\nabla_\mu\nabla_\nu\Phi}{2(n-1)(1-\varepsilon \kappa_n \zeta_c \Phi^2)}.
\end{align}

\subsection{Conformal transformation}
\label{sec:Conformaltr}

We now develop a formulation that enables one to obtain the solutions 
with a conformally coupled scalar field from the solutions with a massless scalar field.
This formulation is insensitive to the particular form of metric {and scalar field ($ \hat g_{\mu\nu},\phi$) in the Einstein frame, and can be presented in two variants. The first variant is focused on the action and the second one on the field equations.}

\subsubsection{Relating the actions} \label{Relatingactions}

For the first approach,
let us assume two metrics, $g_{\mu \nu}$ and $\hat{g}_{\mu \nu}$, conformally related, namely,
\begin{equation} \label{JtoE}
\hat{g}_{\mu \nu}= \omega^2 g_{\mu \nu}.
\end{equation}
The corresponding Ricci scalars are related as 
(see for instance Eq. (2.30) in \cite{Hawking:1973uf})
\begin{equation} \label{sR}
\hat{R}=\omega^{-2} R -2(n-1)\omega^{-3}\square \omega-(n-1)(n-4)\omega^{-4}g^{\mu \nu}\nabla_{\mu}\omega\nabla_{\nu}\omega.
\end{equation}
For $1-\varepsilon\kappa_n \zeta_c \Phi^2>0$, which is always fulfilled for a phantom scalar field, the choice   
\begin{equation} \label{Om}
 \omega= (1-\varepsilon\kappa_n \zeta_c \Phi^2)^{1/(n-2)},
 \end{equation}
is a well-defined function of $\Phi$ for $n>2$,  yielding from \eqref{sR}
\begin{equation} \label{sR2}
\frac{\sqrt{-\hat{g}} \hat{R}}{ 2 \kappa_n}=\sqrt{-g}\left[\frac{(1-\varepsilon\kappa_n \zeta_c \Phi^2)R}{ 2 \kappa_n}+
\frac{\kappa_n \zeta_c \Phi^2}{2(1-\varepsilon\kappa_n \zeta_c \Phi^2)}
g^{\mu \nu}\nabla_{\mu}\Phi\nabla_{\nu}\Phi +\frac{\varepsilon}{4}\square \Phi^2 
\right].
\end{equation}
Using 
$$ \displaystyle \frac{\kappa_n \zeta_c \Phi^2}{1-\varepsilon\kappa_n \zeta_c \Phi^2}=
\frac{\varepsilon}{1-\varepsilon\kappa_n \zeta_c \Phi^2}-\varepsilon,$$
and dropping the last term at the left-hand side of \eqref{sR2} as a total divergence, we obtain
\begin{align} \label{int1}
  \int \D^n x \sqrt{-\hat{g}}\left[\frac{\hat{R}}{2\kappa_n}-\frac{\varepsilon}{2}
\frac{\hat{g}^{\mu \nu}\nabla_{\mu}\Phi\nabla_{\nu}\Phi}{(1-\varepsilon\kappa_n \zeta_c \Phi^2)^2}
\right]=  \int \D^n x \sqrt{-g}\left[\frac{R}{2\kappa_n}-\frac{\varepsilon}{2}\left(
g^{\mu \nu}\nabla_{\mu}\Phi\nabla_{\nu}\Phi+ \zeta_c R \Phi^2\right)\right].
\end{align}
Defining $\phi$ by means of
\begin{equation} \label{Phiphi}
 \frac{\D \Phi}{\D \phi}=\pm (1-\varepsilon\kappa_n \zeta_c \Phi^2),
\end{equation}
the following relation is proved to hold  (up to a boundary term),
\begin{equation}
\int \D^n x \sqrt{-\hat{g}}\left[\frac{\hat{R}}{2\kappa_n}-\frac{\varepsilon}{2}
\hat g^{\mu\nu}\nabla_\mu \phi \nabla_\nu \phi\right]=  
\int \D^n x \sqrt{-g}\left[\frac{R}{2\kappa_n}-\frac{\varepsilon}{2}
(\nabla\Phi)^2- \frac{\varepsilon}{2}\zeta_c R \Phi^2\right].
\label{inter2}
\end{equation}
Thus, Eqs. \eqref{Om} and \eqref{Phiphi} provide the transformation between both frames.

The adequate choice for the case  $1-\varepsilon\kappa_n \zeta_c \Phi^2<0$, which only occurs for $\varepsilon=1$, is 
\begin{equation} \label{Omm}
 \omega= (\kappa_n \zeta_c \Phi^2-1)^{1/(n-2)},
 \end{equation}
Repeating the same steps as before, this choice leads to the following relation between the actions
\begin{equation}
\int \D^n x \sqrt{-\hat{g}}\left[\frac{\hat{R}}{2\kappa_n}-\frac{1}{2}
\hat g^{\mu\nu}\nabla_\mu \phi \nabla_\nu \phi\right]=  -
\int \D^n x \sqrt{-g}\left[\frac{R}{2\kappa_n}-\frac{1}{2}
(\nabla\Phi)^2- \frac{1}{2}\zeta_c R \Phi^2\right],
\label{inter2m}
\end{equation}
where $\phi$ is also defined by \eqref{Phiphi}. Note the overall minus sign in the right hand side of the above equation. This relative minus sign is relevant for the actions, but it does not affect the field equations.

The differential equation \eqref{Phiphi}  crucially depends on $\varepsilon$.  Let us start with the conventional scalar field. In this case, Eq. \eqref{Phiphi} has the following two solutions
\begin{equation}
\sqrt{\kappa_n \zeta_c} \,\Phi = \left\{\begin{array}{c}
\displaystyle \pm \tanh \left(\sqrt{\kappa_n \zeta_c} (\phi-\Phi_0)\right),\\[5mm]
\displaystyle \pm \coth \left(\sqrt{\kappa_n \zeta_c} (\phi-\Phi_0)\right),
 \end{array} \right.
\end{equation}
where $\Phi_0$ is an integration constant.  It is important to note that only one of the above expressions satisfies the required condition ($1-\kappa_n \zeta_c \Phi^2\gtrless 0$). Therefore, for $1-\kappa_n \zeta_c \Phi^2>0$ we have,
\begin{equation} \label{sPhip}
\sqrt{\kappa_n \zeta_c} \,\Phi= 
\pm \tanh \left(\sqrt{\kappa_n \zeta_c} (\phi-\Phi_0)\right),
\end{equation}
and
\begin{equation} \label{sOmegap}
\omega^{-1}= 
\left[\cosh\left( \sqrt{\kappa_n \zeta_c} (\phi-\Phi_0)\right)\right]^{2/(n-2)}.
\end{equation}
For $1-\kappa_n \zeta_c \Phi^2<0$ the proper solution is
\begin{equation}
 \label{sPhipm}
\sqrt{\kappa_n \zeta_c} \,\Phi= 
\pm \coth\left( \sqrt{\kappa_n \zeta_c} (\phi-\Phi_0)\right),
\end{equation}
giving the conformal factor 
\begin{equation} \label{sOmegam}
\omega^{-1}= 
\left[\sinh \left(\sqrt{\kappa_n \zeta_c} (\phi-\Phi_0)\right)\right]^{2/(n-2)}.
\end{equation}

According the second theorem by Bekenstein \cite{Bekenstein:1974sf}, the Einstein-conformal scalar solutions in $n=4$ are obtained in pairs. In fact, this is what also happens in arbitrary dimensions with the pair of solutions  \eqref{sPhip} (labeled as I) and \eqref{sPhipm} (labeled as II), which satisfy $\kappa_n \zeta_c \,\Phi^{\textrm{I}}\Phi^{\textrm{II}}=\pm 1$.

In the phantom case, Eq. \eqref{Phiphi} reads
\begin{equation} \label{Psipsiphantom}
\frac{\D \Phi}{\D \phi}=\pm (1+\kappa_n \zeta_c \Phi^2),
\end{equation}
whose solution is 
\begin{equation} \label{sPsiphantom}
\sqrt{\kappa_n \zeta_c} \,\Phi= 
\pm \tan \left(\sqrt{\kappa_n \zeta_c} (\phi-\Phi_0)\right),
\end{equation}
In this case, the second solution is just obtained by adjusting $\Phi_0$, since $\tan(x+\pi/2)=-\cot(x)=-1/\tan(x)$.

\subsubsection{Using the field equations of both frames} \label{Relatingeom}

As a second independent method, let us focus on the transformation of the equations of motion. 
To this end, we consider the conformal rescaling $g_{\mu\nu}=\Omega^2 \hat g_{\mu\nu}$ 
(note that $\omega=1/\Omega$, where $\omega$ is the conformal factor used in the previous subsection) and 
require that ($ \hat g_{\mu\nu},\phi$) satisfies the Einstein-{massless} scalar field equations
\begin{align}
\label{}
\hat{R}_{\mu\nu}= \kappa_n \epsilon (\hat \nabla_\mu \phi)(\hat \nabla_\nu \phi), \qquad 
\hat{\square} \phi=0.
\end{align}
At this moment, $\epsilon=\pm 1$ should be distinguished from $\varepsilon$. 
Requiring $R=0$ and setting $\Omega=\Omega (\phi)$, the conformal transformation formula (\ref{sR})  leads to 
\begin{align}
\label{}
0=&\Omega^2 R\notag \\
=&\epsilon \kappa_n (\hat \nabla \phi)^2 
-2(n-1)\hat \square \ln \Omega -(n-1)(n-2)(\hat \nabla \ln \Omega)^2
\notag \\
=&\Omega(\phi)^2(\hat\nabla \phi)^2 \left[
\epsilon \kappa_n \Omega(\phi)^2-2(n-1)\Omega(\phi) \Omega''(\phi)-(n-1)(n-4)\Omega'(\phi)^2
\right].
\end{align}
The above equation is integrated to yield
\begin{align}
\label{Omega}
\Omega(\phi)=\left(a_+\cosh({\sqrt{\epsilon \zeta_c \kappa_n} \phi})+\frac{a_-}{\sqrt{\epsilon}}\sinh({\sqrt{\epsilon \zeta_c \kappa_n}\phi})\right)
^{2/(n-2)},
\end{align}
where $a_\pm$ are real constants. Inserting this into
Einstein's equations, we have three terms proportional to $\hat g_{\mu\nu}$, 
$\hat \nabla_\mu\phi \hat \nabla_\nu \phi$, 
and $\hat \nabla_\mu \hat \nabla_\nu\phi$. The terms proportional to 
$\hat \nabla_\mu \hat \nabla_\nu\phi$ determines $\Phi(\phi)$ up to an integration constant, 
which can be fixed by the rest of Einstein's equations to be 
\begin{align}
\label{CS}
\Phi(\phi)=\pm \frac {\sqrt{\epsilon}}{\sqrt{\zeta_c \varepsilon\kappa_n}}
\frac{(a_-/\epsilon)\cosh(\sqrt{\epsilon \zeta_c \kappa_n} \phi)+(a_+/\sqrt{\epsilon})
\sinh(\sqrt{\epsilon \zeta_c \kappa_n} \phi)}{a_+\cosh(\sqrt{\epsilon \zeta_c\kappa_n } \phi)+(a_-/\sqrt{\epsilon})
\sinh(\sqrt{\epsilon \zeta_c \kappa_n}\phi)}.
\end{align}
Then we must have $\varepsilon=\epsilon$ for the scalar field $\Phi$ to be real. Note that the transformation given by \eqref{Omega}-\eqref{CS} is effectively generated by just one parameter, namely $a_-/a_+$, since $\Omega$ is defined up to an overall factor.
The following analysis will be split into two cases according to $\epsilon=1$ or $\epsilon=-1$.

Let us first consider the non-phantom case ($\epsilon=1$).
The solution is specified by a single parameter $a_-/a_+$, so that 
we have thus three cases to consider: 
(i) $|a_-/a_+|<1$, (ii) $|a_-/a_+|>1$,  (iii) $|a_-/a_+|=1$. 
In the case (iii), equation (\ref{CS}) implies that the conformal scalar field is {$\Phi=\pm (\sqrt{\zeta_c \kappa_n})^{-1}$}.   
As mentioned before, we shall not explore this case by virtue of $1-\zeta_c\kappa_n \Phi^2=0$, for which 
the coefficient of the scalar curvature in the action (\ref{actionconf}) vanishes on shell. 
This case corresponds to the strong coupling, at which the theory is not predictable, reducing just to $R=0$. 
The cases (i) and (ii) should also be distinguished, because the coefficient in front of the scalar curvature 
in the action has opposite sign $1-\zeta_c\kappa_n \Phi^2\gtrless 0$. Since the overall factor of $\Omega$ in (\ref{Omega}) is irrelevant, 
we can set $a_-/a_+=-\tanh( \sqrt{\zeta_c\kappa_n}\Phi_0)$ when $|a_-/a_+|<1$, where  $\Phi_0$ is a constant. 
We thus obtain the transformation
\begin{align}
\label{conformalep1case1}
\Omega =\left[\cosh\left(\sqrt{\zeta_c\kappa_n}(\phi-{\Phi_0})\right)\right]^{2/(n-2)}, \qquad 
\Phi= \pm \frac{1}{\sqrt{\zeta_c\kappa_n}}\tanh \left(\sqrt{\zeta_c\kappa_n} (\phi-{\Phi_0})\right),
\end{align}
{for the case (i)}. In the case of $|a_-/a_+|>1$, we can set {$a_-/a_+=-1/\tanh( \sqrt{\zeta_c\kappa_n}\Phi_0)$} with $\Phi_0$ being a constant, for which the transformation is given by
\begin{align}
\label{conformalep1case2}
\Omega =\left[\sinh\left(\sqrt{\zeta_c\kappa_n}(\phi-{\Phi_0})\right)\right]^{2/(n-2)},  \qquad 
\Phi= \pm \frac{1}{\sqrt{\zeta_c\kappa_n}}\coth \left(\sqrt{\zeta_c\kappa_n}(\phi-{\Phi_0})\right).
\end{align}

Let us next consider the phantom case $\epsilon=-1$. Ignoring the overall factor, 
one can set $a_-/a_+=\tan( \sqrt{\zeta_c\kappa_n}\Phi_0)$, 
thereby 
\begin{align}
\label{conformalepm1}
\Omega(\phi)=\left[\cos \left(\sqrt{\zeta_c\kappa _n}(\phi-{\Phi_0})\right)\right]^{2/(n-2)},  \qquad 
\Phi=\pm\frac{1}{\sqrt{\zeta_c\kappa _n}} \tan \left(\sqrt{\zeta_c\kappa _n} (\phi-{\Phi_0})\right).
\end{align}
Contrary to the non-phantom case, the conformal factor is a trigonometric function. 
This alters the causal structure of the solution  considerably, since 
$\sqrt{\zeta_c\kappa _n}(\phi-{\Phi_0})=\pm \pi/2$ corresponds to the 
curvature singularity where the square of the Weyl tensor (\ref{Weylsq}) 
necessarily blows up. 

Note that this approach provides the same transformations presented in the previous section \ref{Relatingactions}. Thus, the two variants employed here to obtain the transformation relating both frames are equivalent. In the following, we will obtain the explicit form of metric with a conformally coupled scalar field, by taking the solutions obtained in section \ref{sec:classification} as a seed.

\subsection{All possible solutions} 

Exploiting  the formulation developed in \ref{sec:Conformaltr}, 
we construct all possible solutions for (\ref{EOMcc}) with the 
symmetry (\ref{gauge-higher}), by taking the solutions derived in section \ref{sec:classification} as seed solutions.  We only classify solutions and do not attempt to go into the detail of the physical and causal properties of the solutions, since the analysis of each solution requires laborious works more than that in the Einstein frame. We leave this issue to the future work. {We start with the conventional (non-phantom) conformal scalar field configurations.}

\subsubsection{Non-phantom Fisher solution: branch I} 
\label{|a_-/a_+|<1}

Let us take the Fisher solution (\ref{JNWk}) as a seed ($\hat M, \hat g_{\mu\nu}$) and perform the transformation \eqref{conformalep1case1}. Introducing\footnote{Although the transformation has a free parameter $\Phi_0$, this is added to the arbitrary constant $\phi_0$ yielding a general solution with three integration constants, which is the same number within the seed general solution.} $f_0:=\exp[\sqrt{\zeta_c\kappa_n}(\Phi_0-{\phi_0})/\gamma]$,
we have 
\begin{align}
\label{confJNW1}
\D s^2=&\left[\frac{(f/f_0)^\gamma+(f/f_0)^{-\gamma}}{{2}}\right]^{4/(n-2)}\left[
-f^\alpha \D t^2+f^{-(\alpha+n-4)/(n-3)}(\D r^2+r^2 f \D \Sigma_{k,n-2}^2)
\right],  \\
\Phi=& \pm\frac{1}{\sqrt{\zeta_c \kappa_n}} 
\frac{(f/f_0)^{2\gamma}-1}{(f/f_0)^{2\gamma}+1}, 
\end{align}
where $f=|k-M/r^{n-3}|$ is non-negative for ensuring a real solution and 
\begin{align}
\label{gamma}
\gamma := {\pm} \frac{n-2}{4}\sqrt{\frac{\epsilon(1-\alpha^2)}{(n-1)(n-3)}},
\end{align}
with $\epsilon=+1$. 
When $n=4$ and $\alpha=1/2$, this solution recovers the one given in 
\cite{Bocharova:1970skc,Bekenstein:1974sf} and admits a degenerate horizon. 
{In the spherically symmetric higher-dimensional case ($K^{n-2}=S^{n-2}$)}, there exist no analogous solutions possessing a regular horizon \cite{Xanthopoulos:1992fm,Klimcik:1993cia}.

\subsubsection{Non-phantom Fisher solution: branch II}

Taking the Fisher solution as a seed and considering the transformation \eqref{conformalep1case2}, we have 
\begin{align}
\label{confJNW2}
\D s^2=& \left[\frac{(f/f_0)^\gamma-(f/f_0)^{-\gamma}}{2}\right]^{4/(n-2)}\left[
-f^\alpha \D t^2+f^{-(\alpha+n-4)/(n-3)}(\D r^2+r^2 f \D \Sigma_{k,n-2}^2)
\right], \\
\Phi=& \pm \frac{1}{\sqrt{\zeta_c\kappa_n}}\frac{(f/f_0)^{2\gamma}+1}{(f/f_0)^{2\gamma}-1},
\end{align}
where $f, f_0$ and $\gamma$ are the same as in the previous section \ref{|a_-/a_+|<1}.
Compared to the branch {I} solution (\ref{confJNW1}), the conformal scalar field  is related by 
the inverse relation $(\sqrt{\zeta_c\kappa_n}\Phi) \to {\pm} (\sqrt{\zeta_c\kappa_n}\Phi)^{-1}$ {as announced before.}

To classify the phantom solutions, we consider the transformation (\ref{conformalepm1}) in the remaining of this subsection.

\subsubsection{Phantom Fisher solution}

Taking the phantom Fisher solution (\ref{JNWk}) as a seed, the 
solution is given by 
\begin{align}
\label{confJNWphantom}
\D s^2=& \left[\cos (\gamma\ln ({f/f_0}))\right]^{4/(n-2)}\left[
-f^\alpha \D t^2+f^{-(\alpha+n-4)/(n-3)}(\D r^2+r^2{f} \D \Sigma_{k,n-2}^2)
\right], \\
\Phi=& \pm \frac{1}{\sqrt{\zeta_c\kappa_n}}\tan (\gamma \ln{f/f_0}),
\end{align}
where $\gamma$ is understood to be (\ref{gamma}) with $\epsilon=-1$.

\subsubsection{Gibbons solution}

Let us take the Gibbons solution (\ref{multiGibbons}) 
as a seed metric. Setting $\epsilon=-1$, the solution in the conformal frame reads 
\begin{align}
\label{confGibbons}
\D s^2=& \left[\cos (\gamma' (H-H_0))\right]^{4/(n-2)}\left[
-e^{-H} \D t^2+e^{H/(n-3)}h_{IJ}\D x^I \D x^J 
\right], \\
\Phi=& \pm \frac{1}{\sqrt{\zeta_c\kappa_n}}\tan (\gamma' (H-H_0)),
\end{align}
where $H_0$ is an arbitrary constant, $\Delta_h H=0 $, $h_{IJ}$ is a Ricci-flat metric and 
\begin{align}
\label{gammap}
\gamma' := {\pm} \frac{n-2}{4}\sqrt{\frac{1}{(n-1)(n-3)}}.
\end{align}

\subsubsection{Ellis-Bronnikov solution}

Lastly, let us consider the Ellis-Bronnikov solution (\ref{Ellis-Bronnikov})  as a seed. 
Setting $\epsilon=-1$, the conformal frame metric is 
\begin{align}
\label{confEliis-Bronnikov}
\D s^2=& \left[\cos (\gamma'' (U-{U_0}))\right]^{4/(n-2)}\left[
-e^{-2\beta U} \D t^2+e^{2\beta U/(n-3)}V^{1/(n-3)}\left(\frac{\D r^2}{V}+r^2 \D \Sigma_{k=1,n-2}^2\right)
\right],  \\
\Phi=& \pm \frac{1}{\sqrt{\zeta_c\kappa_n}}\tan (\gamma'' (U-{U_0})),
\end{align}
where $U$, $V$ are given by (\ref{Ellis-BronnikovUV}), {$U_0$ is an arbitrary constant} and  
\begin{align}
\label{gammapp}
\gamma'' := {\pm}\frac{n-2}{2}\sqrt{\frac{(1+\beta^2)}{(n-1)(n-3)}}.
\end{align}

\subsection{Consistency check: solving field equations in the Jordan frame}

In this section we explore if the conformal transformation from 
the Einstein-massless scalar system enables us to obtain all possible solutions. 
The best way to address the question is to solve directly the field equations (\ref{EOMcc}).
For the sake of clarity to readers, we show that this is indeed a viable task and that both methods provide the same results.

We consider the same class of metrics as in Sec. \ref{sec:classification}, but with a bit different radial coordinate $\rho$ defined as $ \D \rho= G^{-1} \D x $. Then, the general metric \eqref{gauge-higher} reads 
\begin{align}
\D s^2=&-F(\rho)^{-2}\D t^2+F(\rho)^{2/(n-3)}G(\rho)^{1/(n-3)}\biggl(G(\rho)\D \rho^2+\gamma_{ab}(z)\D z^i\D z^j\biggl),\label{gauge-higher2}
\end{align}
and the conformal scalar field is assumed to be a function of the new radial coordinate, $\Phi=\Phi(\rho)$. Using the gauge (\ref{gauge-higher2}), and defining 
\begin{equation} \label{rede}
F:=e^{-b} \quad  \mbox{and} \quad  G:= h^{-2},
\end{equation}
we obtain  from $E^{t}{} _{t}, E^{r} {}_{r}, E^{i} {}_{j}$ and $\square \Phi=0$, the following set of equations
\begin{align}
&b''-\frac{2 \varepsilon \kappa_n \zeta_c   \left( \Phi '^2+(n-2)b' \Phi  \Phi '\right)}{(n-2) \left(1-\varepsilon \kappa_n \zeta_c \Phi^2\right)}= 0,  
\label{tt0}\\
& h''+\frac{h\left(b''-(n-2) b'^2\right)}{n-2} + \frac{4  \varepsilon\kappa_n \zeta_c \Phi  \left(h b' \Phi '+(n-2) h' \Phi '+(n-3) h \Phi ''\right)-\varepsilon\kappa_n  (n-3) h \Phi '^2}{2(n-2) \left(1- \varepsilon\kappa_n \zeta_c  \Phi^2\right)}= 0, \label{rr0}\\
&\left\{h''+\frac{h^2 b''-h'^2+k(n-3)^2}{h}-\frac{2 \varepsilon \kappa_n \zeta_c  \left((n-2) \Phi \Phi '\left(h b'+h'\right)-(n-3) h \Phi '^2\right)}{(n-2) \left(1-\varepsilon \kappa_n  \zeta_c \Phi^2\right)}\right\}\delta^{i}{}_{j}  = 0, \label{ab0}\\
&\Phi ''=0, \label{KG0}
\end{align}
where a prime denotes the derivative with respect to $\rho$.  In these coordinates, the equation for the conformal scalar field is easily integrated as 
\begin{align}
\Phi=\Phi_1 \rho + \Phi_0,
\end{align}
where $\Phi_0$ and $\Phi_1$ are integration constants. Hereafter, we consider  $\Phi_1 \neq 0$ for avoiding a constant scalar field. As a second advantage of the present coordinate system, one can observe that Eq. \eqref{tt0} is a linear differential equation for $b'$. Once $b$ is determined, Eq. \eqref{rr0} is also linear for $h$. Thus, Eq.  \eqref{ab0} becomes an algebraic constraint for the integration constants appearing in $b$ and $h$. The fact we are dealing with a set of linear differential equations allows us to find the general solution. Additionally, since we are considering a non-constant scalar field, we can use the scalar field as a coordinate by means of the substitution  $\rho= (\Phi-\Phi_0)/\Phi_1$, so that the general solution can be expressed in terms of $\Phi$.

The integration of \eqref{tt0} is straightforwardly done as
\begin{equation} \label{solett}
 b-b_0=\left\{ \begin{array}{ll}
\displaystyle 2b_1 \tanh ^{-1}\left(\sqrt{\kappa_n \zeta_c  } \Phi \right)-\frac{\ln\left(1- \kappa_n \zeta_c  \Phi ^2\right)}{n-2},& \varepsilon=1, \:\kappa_n \zeta_c  \Phi ^2<1,\\[4mm]
\displaystyle 2b_1 \coth^{-1}\left(\sqrt{\kappa_n\zeta  } \Phi \right)-\frac{\ln\left( \kappa_n \zeta_c  \Phi ^2-1\right)}{n-2},
       & \varepsilon=1, \: \kappa_n \zeta_c  \Phi ^2>1,\\[4mm] 
\displaystyle 2b_1 \tan^{-1}\left(\sqrt{\kappa_n \zeta_c} \Phi \right)-\frac{\ln\left(1+  \kappa_n \zeta_c \Phi ^2\right)}{n-2}, & \varepsilon=-1, 
   \end{array} \right.
\end{equation}
where $b_0, b_1$ are integration constants. Replacing  \eqref{solett} into \eqref{rr0} yields the linear equation
\begin{equation}
\left(1- \varepsilon \kappa_n \zeta_c  \Phi ^2\right) h''(\Phi )+2 \varepsilon \kappa_n \zeta_c   \Phi  h'(\Phi )-2 \varepsilon \kappa_n \zeta_c \left(\frac{2 \varepsilon b_1^2 (n-2)^2-2}{(n-2)^2 (1-\varepsilon\kappa_n \zeta_c   \Phi ^2)}+1\right) h(\Phi ) =0,
\end{equation}
which leads to five different solutions according to $\varepsilon$ and $\Phi$, enumerated below with Roman numbers,  
\begin{equation} \label{solerr}
 h=\left\{ \begin{array}{lcl}
 \left(1- \kappa_n \zeta_c  \Phi ^2\right) \left(h_0 e^{2\sqrt{a_+} x_{<}}+h_1 e^{-2\sqrt{a_+} x_{<}}\right),& \text{I:} & \varepsilon=1, \kappa_n \zeta_c  \Phi ^2<1,\\[4mm]
 \left( \kappa_n \zeta_c  \Phi ^2-1\right) \left(h_0 e^{2\sqrt{a_+} x_{>}}+h_1 e^{-2\sqrt{a_+} x_{>}}\right),&\text{II:} &  \varepsilon=1, \kappa_n \zeta_c  \Phi ^2>1,\\[4mm] 
\left(1+ \kappa_n \zeta_c  \Phi ^2\right) \left( h_0 \cosh \left(2 \sqrt{a_-} y\right)+h_1 \sinh \left(2 \sqrt{a_-} y\right)\right),& \text{III:} &\varepsilon=-1,  a_- > 0,\\[4mm] 
\left(1+ \kappa_n \zeta_c  \Phi ^2\right) \left(h_0 \cos \left(2 \sqrt{-a_-} y\right)+h_1 \sin \left(2 \sqrt{-a_-} y\right) \right),& \text{IV:} &\varepsilon=-1,  a_- < 0,\\[4mm] 
\left(1+ \kappa_n \zeta_c  \Phi ^2\right) \left(h_0 +h_1 y\right), &\text{V:}  & \varepsilon=-1,  a_- =0,
   \end{array} \right.
\end{equation}
where $h_0, h_1$ are integration constants and
\begin{equation}
x_<:=\tanh ^{-1}\left(\sqrt{\kappa_n \zeta_c  } \Phi \right), \quad x_>:=\coth ^{-1}\left(\sqrt{\kappa_n \zeta_c  } \Phi \right), \quad y:= \tan ^{-1}\left(\sqrt{\kappa_n \zeta_c} \Phi \right),
\end{equation}
with
\begin{equation}
a_{\pm}=\frac{ (n-2)^2 b_1^2\pm (n-3) (n-1)}{(n-2)^2}.
\end{equation}

Replacing \eqref{solett} and \eqref{solerr} in \eqref{ab0}, we obtain  algebraic relations between the integration constants and the curvature $k$ of the Einstein space $K^{n-2}$,
\begin{equation} \label{soleij}
\begin{array}{cl}
\text{I, II:} &\displaystyle 0= 16 a_+ \, h_0 h_1 \kappa_n \zeta_c  \Phi_1 ^2 +k (n-3)^2, \\[4mm]
\text{III:} &\displaystyle 0= 4 a_- \, \kappa_n \zeta_c \Phi_1 ^2 \left(h_0^2-h_1^2\right)+k (n-3)^2, \\[4mm] 
\text{IV:} &\displaystyle 0= 4 a_- \, \kappa_n \zeta_c \Phi_1 ^2 \left(h_0^2+h_1^2\right)+k (n-3)^2, \\[4mm] 
\text{V:} & \displaystyle 0= h_1^2 \kappa_n  \zeta_c \Phi_1 ^2-k (n-3)^2.
\end{array}
\end{equation}
The above equations fix the amplitude of the constant $\Phi_1$ appearing in the scalar field in terms of the constants appearing in the metric, $b_1, h_0$, and  $h_1$.\footnote{A rescaling of the $t$ coordinate and the addition of a global constant factor allow us to set $b_0=0$.} We note that phantom solutions IV and V exist only for $k=1$, while I, II, and III allow $k=0,\pm1$. 

To compare with the conformal transformations exhibited in the previous sections, the general solution can be written in the form 
\begin{align} \label{gensol2}
\D s^2&=W^{-2/{(n-2)}}\left[ -f_0 \D t^2+\left[f_0^{-1} f_1^{-2(n-2)}\right]^{1/(n-3)}\biggl(\D x^2+f_1^2\gamma_{ij}(z)\D z^i\D z^j\biggl)\right], \\
&:=W^{-2/{(n-2)}} \D s_{\text{E}}^2,
\end{align}
with 
\begin{align} \label{gensol2p}
x=\int W^{-1} \D \rho, \qquad f_0=\exp \left[ 4b_1 \sqrt{\kappa_n \zeta_c}\phi \right], \qquad  \phi=\Phi_1 (x -x_0),
\end{align}
where the remaining functions $W$, $\Phi$, and $f_1$ are displayed in Table \ref{tablegensol}.

\begin{table}[t]
\centering
\caption{General solution for a massless conformally coupled scalar field and correspondence between Jordan (\textbf{J}) and Einstein (\textbf{E}) frames. Note that $W^{-2/{(n-2)}}=\Omega^2$.}  \label{tablegensol}
\vskip 1mm
{\small 
\begin{tabular}{|c|c|c|c|c|}
\hline 
\textbf{J} & $W$ & $\sqrt{\kappa_n \zeta_c  } \Phi$ & $f_{1}$ &\textbf{E}\\
\hline
I & $1- \kappa_n \zeta_c  \Phi ^2$ &  $ \tanh \left(\sqrt{\kappa_n \zeta_c  } \phi \right)$   & $h_0 e^{2\sqrt{a_+} \sqrt{\kappa_n \zeta_c} \phi }+h_1 e^{-2\sqrt{a_+}\sqrt{\kappa_n \zeta_c} \phi }$ & Fisher, branch 1 \\
\hline 
II & $\kappa_n \zeta_c  \Phi ^2-1$ &  $ \coth \left(\sqrt{\kappa_n \zeta_c  } \phi \right)$    & $h_0 e^{2\sqrt{a_+} \sqrt{\kappa_n \zeta_c} \phi }+h_1 e^{-2\sqrt{a_+} \sqrt{\kappa_n \zeta_c} \phi} $ & Fisher, branch 2  \\
\hline 
III & $1+ \kappa_n \zeta_c  \Phi ^2$ &  $\tan\left(\sqrt{\kappa_n \zeta_c} \phi \right)$     & $h_0 \cosh \left(2 \sqrt{a_-} \sqrt{\kappa_n \zeta_c} \phi \right)+h_1 \sinh \left(2 \sqrt{a_-} \sqrt{\kappa_n \zeta_c} \phi \right)$ & phantom Fisher  \\
\hline 
IV & $1+\kappa_n \zeta_c  \Phi ^2$ &  $\tan\left(\sqrt{\kappa_n \zeta_c} \phi \right)$    & $h_0 \cos \left(2 \sqrt{-a_-} \sqrt{\kappa_n \zeta_c} \phi \right)+h_1 \sin \left(2 \sqrt{-a_-} \sqrt{\kappa_n \zeta_c} \phi \right) $ & Ellis-Bronnikov \\
\hline 
V & $1+ \kappa_n \zeta_c  \Phi ^2$ &  $\tan\left(\sqrt{\kappa_n \zeta_c} \phi \right)$    & $h_0 +h_1 \sqrt{\kappa_n \zeta_c} \phi $ & Ellis-Gibbons\\ 
\hline 
\end{tabular} }
\end{table}

Using the radial coordinate $ \int f_1^{-2} \D x$ and the identifications  $G=f_1^{-2}$ and $F^{-2}=f_0$, we note that $\D s_{\text{E}}^2$ and $\phi$ exhibited in the general solution correspond to the line element and massless minimally coupled scalar field obtained in Sec. \ref{sec:classification}. Additionally, the conformal factor and conformal scalar field displayed in Table \ref{tablegensol} match those given for the conformal transformation previously discussed in this section. Thus, the general solution obtained by a direct integration of the field equations in the Jordan frame is equivalent to the one achieved by a conformal transformation applied to the general solution in the Einstein frame. Table \ref{tablegensol} summarizes the correspondence between both frames.

\section{Summary and future prospects}
\label{sec:conclusion}

In this paper, we have studied various aspects of the $n$-dimensional static solutions 
sourced by a (non-)phantom scalar field. We have generalized classification program by Ellis \cite{Ellis1973} for the four dimensional spherical case into higher dimensions with the sphere part replaced by an arbitrary Einstein space. We have found  the generalized Fisher solution (\ref{JNWk}) (valid for $k=0, \pm 1$), the generalized Ellis-Gibbons solution  (\ref{Gibbonsol}), the generalized Ellis-Bronnikov solution (\ref{Ellis-Bronnikov})  and the plane-symmetric solution (\ref{Buchdahl}). We have uncovered a unified description of metrics (\ref{Gibbonsol}) and  (\ref{Buchdahl}) in the framework of the Gibbons solutions. Since the non-phantom field only admits a Fisher class, the diversity of solutions in this class of metrics is a marvelous feature of a phantom field. We summarize the results in Table (\ref{table:solutions}).

\begin{table}[htb]
\begin{center}
\caption{\label{table:solutions} A complete list of static solutions with pseudo-spherical symmetry.}
\vspace{0.2cm}
{\small 
\begin{tabular}{|c|c|c|c|c|}
\hline
frame& scalar &$k=1$ &$k=-1$ & $k=0$ \\ 
\hline\hline
\multirow{2}{*}{Einstein} &non-phantom &  \multicolumn{3}{|c|}{Fisher (\ref{JNWk})}  \\
\cline{2-5}
&phantom & 
\begin{tabular}{c}Fisher (\ref{JNWk}), Ellis-Gibbons (\ref{Gibbonsol}),\\
 Ellis-Bronnikov (\ref{Ellis-Bronnikov}) \end{tabular} 
 & Fisher (\ref{JNWk}) & \begin{tabular}{c} Fisher  (\ref{JNWk}), \\ Gibbons (\ref{Buchdahl})\end{tabular}
\\
\hline
\multirow{2}{*}{Conformal} &non-phantom &  \multicolumn{3}{|c|}{Fisher (\ref{confJNW1}), (\ref{confJNW2})}  \\
\cline{2-5}
&phantom & 
\begin{tabular}{c}
Fisher (\ref{confJNWphantom}), Ellis-Gibbons (\ref{confGibbons}),\\
 Ellis-Bronnikov (\ref{confEliis-Bronnikov}) \end{tabular} 
 & Fisher (\ref{confJNWphantom}) & \begin{tabular}{c}Fisher  (\ref{confJNWphantom}),\\
  Gibbons (\ref{confGibbons}) \end{tabular}
\\
\hline
\end{tabular} 
}
\end{center}
\end{table} 

After the encyclopedic enumeration of the solutions, we have spelled out the physical properties of the solutions. Our intriguing revelation is that the Fisher solution and the Ellis-Gibbons solution enjoy 
naked p.p curvature singularities in a parameter region where there are no scalar curvature singularities. 
This leads us to conclude that these geometries do not describe regular wormhole solutions. 
What was curious to us is that  the locus of these p.p curvature singularities corresponds to the infinite areal radius,  but is reachable within a finite affine time for radial null geodesics. From the present analysis, it is not certain to us whether this peculiar behavior only occurs in a spacetime which violates energy conditions. Nevertheless, we deduce that the existence of an area-diverging surface  which is achievable within a finite affine time is a strong indication for the appearance of p.p curvature singularity. To illustrate, let us consider the flat Friedman-Lema\^itre-Robertson-Walker universe with a phantom equation of state $p =w \rho$  ($-5/3<w<-1$), which admits this kind of surface (F4a of Table 1 in \cite{Harada:2018ikn}). One can easily show that this surface corresponds to the p.p curvature singularity, supporting positively our conjecture.  Regardless of the soundness for this speculation, an important lesson from our results is that the areal radius is not a definitive measure to examine the spacetime structure.

Only a wormhole candidate is therefore the Ellis-Bronnikov solution, which we have investigated in detail the throat structure. Since many of the past works have been limited to the zero mass wormholes, it has not been fully recognized that the throat does not always correspond to the minimal radius and that the gluing procedure is more sensitive in higher dimensions. Nevertheless, we have clarified that Ellis-Bronnikov solution with a nonvanishing {mass} also describes a genuine wormhole with two-sided asymptotically flat regions.
To examine the stability for the Ellis-Bronnikov solution in arbitrary dimensions by generalizing the work of \cite{Torii:2013xba,Blazquez-Salcedo:2018ipc} is definitely an interesting subject to be studied in the future.

We also presented the classification of the static solutions in the conformal frame. First, we introduced (by means of two approaches, proved to be equivalent) a simple transformation formula which allows us to obtain the 
solution with a conformally coupled scalar field from the solution with a 
massless scalar field. Second, we get the same results by solving the field equations in the Jordan frame. As shown in Table \ref{table:solutions}, the system with a conformally
coupled scalar inherits a broader variety of solutions.  The physical and causal properties of these solutions will be an interesting future work to be investigated. We also leave the full classification of solutions in Jordan frame where the coupling constant $\zeta$ is arbitrary to a future study.  

By taking the three family of solutions obtained in this paper as a seed, 
it is possible to obtain charged solutions by using the symmetry of the target space for the static system. 
The (anti-)de Sitter generalization can also be done by introducing the scalar potential. 
These issues are reported in forthcoming papers \cite{paperII,paperIII}. 
It would also be of considerable  physical interest to construct rotating solutions along the line of \cite{Bogush:2020lkp}.

\subsection*{Acknowledgements}

We thank Hideki Maeda for valuable discussions and useful comments to our manuscript. 
M. N thanks Hokkai-Gakuen University for a kind hospitality, where a part of this work was carried out. 
The work of M. N is partially supported by 
Grant-in-Aid for Scientific Research (A) from JSPS 17H01091 and (C) 20K03929.
The work of C. M has been partially funded by Fondecyt grants  1201208 and 1180368. The Centro de Estudios Cient\'{\i}ficos (CECs) is funded by the Chilean Government through the Centers of Excellence Base Financing Program of Conicyt.

\appendix

\section{Geometric quantities}
\label{app:geom}

Suppose the $n$-dimensional spacetime 
to be a warped product of an 
$(n-2)$-dimensional {Einstein} space $(K^{n-2}, \gamma _{ij})$
and a two-dimensional spacetime $(M^2, g_{AB})$. Namely, the line element
is given by
\begin{align}
g_{\mu \nu }\D x^\mu \D x^\nu =g_{AB}(y)\D y^A\D y^B +S^2(y) \gamma _{ij}(z)\D z^i\D z^j,
\label{eq:ansatz}
\end{align} 
where
$A,B = 0, 1;~i,j = 2, ..., n-1$. 
Here $S$ is a scalar on $(M^2, g_{AB})$ and $\gamma_{ij}$ is the metric on {the Einstein space $K^{n-2}$, whose  Ricci tensor is normalized as ${}^{(\gamma)}R_{ij}=(n-3)k \gamma_{ij}$, with} $k = \pm 1, 0$. 
The non-vanishing components of the Levi-Civita connections are
\begin{align}
\begin{aligned}
{\Gamma ^A}_{BC}&={}^{(2)}{\Gamma ^A}_{BC }(y),\quad 
{\Gamma ^i}_{jk}={\hat{\Gamma} ^i}_{~jk}(z), \\
{\Gamma ^A}_{jk}&=-S(D^A S) \gamma _{jk},\quad 
{\Gamma ^i}_{jA}=\frac{D_A S}{S}{\delta ^i}_j, 
\end{aligned}
\end{align}
where the superscript (2) denotes the two-dimensional quantity and $D_A$ is the two-dimensional covariant derivative compatible with
$g_{AB}$.
The non-vanishing components of the Riemann tensors are
\begin{subequations}
\label{eq:Riemann}
\begin{align}
{{R}^A}_{BCD}&={}^{(2)}{{R}^A}_{BCD},\\
{{R}^A}_{iBj}&=-S(D^A D_B S)\gamma _{ij},
\\
{{R}^i}_{jkl}&={}^{(\gamma)}R^i{}_{jkl}-(DS)^2({\delta ^i}_k\gamma _{jl}-{\delta ^i}_l\gamma _{jk}),
\end{align}
\end{subequations}
where $(DS)^2:=(D_AS)(D^AS)$ and ${}^{(\gamma)}R^i{}_{jkl}$ is the Riemann tensor constructed out of $\gamma_{ij}$.
The Ricci tensor and the Ricci scalar are given by 
\begin{subequations}
\label{eq:Ricci}
\begin{align}
{R}_{AB}&={}^{(2)}{R}_{AB}-(n-2)\frac{D_AD_BS}{S},\\
{R}_{ij}&=\left\{-S D^2S+(n-3)[k-(DS)^2]\right\}\gamma _{ij},  \\
{R}&={}^{(2)}{R}-2(n-2)\frac{D^2S}{S}+(n-2)(n-3)\frac{k-(DS)^2}{S^2}, 
\end{align}
\end{subequations}
where $D^2S:=D_AD^AS$ and ${}^{(\gamma)}R_{ij}=(n-3)k \gamma_{ij}$ was used.
The Kretschmann scalar ${K}:={R}^{\mu\nu\rho\sigma}{R}_{\mu\nu\rho\sigma}$ is 
\begin{align}
{K}=
&{}^{(2)}{R}^2 
+4(n-2)\frac{(D_AD_BS)(D^AD^BS)}{S^2}+2(n-2)(n-3)\frac{(k-(DS)^2)^2}{S^4}\notag\\
&+\frac1{S^4}\left[{}^{(\gamma)}R_{ijkl}{}^{(\gamma)}R^{ijkl} -2k^2(n-2)(n-3)\right].
\label{Kre}
\end{align}
When $\gamma_{ij}$ is a metric of a constant curvature space, we have 
${}^{(\gamma)}R_{ijkl}=2k\gamma_{i[k}\gamma_{l]j}$, implying that 
the last term in (\ref{Kre}) drops off.

The Weyl tensor is given by (see e.g, \cite{Maeda:2007uu})
\begin{subequations}
\begin{align}
\label{}
C_{ABCD}=&\frac{n-3}{n-1}W g_{A[C}g_{D]B}, \\
C_{AiBj}=&-\frac{n-3}{2(n-1)(n-2)}W g_{AB} S^2 \gamma_{ij},  \\
C_{ijkl}=& {}^{(\gamma)} C_{ijkl}S^2+\frac{2 W}{(n-1)(n-2)}S^4 \gamma_{i[k}\gamma_{j]l},
\end{align}
\end{subequations}
where
\begin{align}
\label{W}
W:={}^{(2)}R+2\frac{D^2 S}{S}+2\frac{k-(DS)^2}{S^2}.
\end{align}
The Weyl square ${C^2}:=C_{\mu\nu\rho\sigma}C^{\mu\nu\rho\sigma}$ 
reads 
\begin{align}
\label{Weylsq}
{C^2}=\frac 1{S^4}{}^{(\gamma)} C_{ijkl}{}^{(\gamma)} C^{ijkl}
+\frac{n-3}{n-1}W^2.
\end{align}
When $\gamma_{ij}$ is a metric of a constant curvature space, the first term of the right-hand
side of this equation vanishes.

In the main text, we meet static spacetimes with the following form
\begin{align}
\label{metricf1f2S}
\D s^2=-f_1(r)\D t^2+f_2(r)\D r^2+S^2(r) \gamma_{ij}(z)\D z^i \D z^j, 
\end{align}
where $f_1$ and $f_2$ are functions only of $r$. 
Then, the tangent vector $k^\mu $ for an affinely parameterized outgoing radial null geodesics 
is given by 
\begin{align}
\label{}
k^\mu = \frac{1}{f_1(r)} \left(\frac{\partial}{\partial t} \right)^\mu +\frac{1}{\sqrt{f_1(r)f_2(r)}}
\left(\frac{\partial}{\partial r} \right)^\mu, \qquad 
k^\nu \nabla_\nu k^\mu=0.
\end{align}
One can thus constitute the pseudo-orthonormal frame
\begin{align}
\label{}
g_{\mu\nu}=-2k_{(\mu }n_{\nu)} +\delta_{\hat i\hat j}E^{\hat i}{}_\mu E^{\hat j}{}_\nu,  
\end{align}
by 
\begin{align}
\label{nEi}
n^\mu =\frac 12 \left(\frac{\partial}{\partial t} \right)^\mu 
-\frac{\sqrt{f_1(r)}}{2\sqrt{f_2(r)}}\left(\frac{\partial}{\partial r} \right)^\mu, \qquad 
E_{\hat i}{}^\mu = \frac 1{S(r)} e_{\hat i}{}^{i}\left(\frac{\partial}{\partial z^i}\right)^\mu,
\qquad 
\gamma^{ij}=\delta^{\hat i\hat j}e_{\hat i}{}^{i}e_{\hat j}{}^{j},
\end{align}
where $\hat i, \hat j=1,..,n-2$. It is elementary to verify that 
these bases are parallelly propagated along $k^\mu$
\begin{align}
\label{}
k^\nu \nabla_\nu n^\mu=k^\nu \nabla_\nu E_{\hat i}{}^\mu=0.
\end{align}
In this frame, the components of the Riemann tensor relevant to 
the p.p curvature singularity are 
\begin{align}
\label{Rili}
R_{\mu\nu\rho\sigma}k^\mu E_{\hat i}{}^\nu k^\rho E_{\hat j}{}^\sigma 
&=-S^{-1}k^A k^B (D_A D_B S)  \delta_{\hat i\hat j}
=\frac{f_2f_1'S'+f_1 f_2'S'-2 f_1 f_2 S''}{2f_1^2 f_2 ^2 S}\delta_{\hat i\hat j},\\
\label{Rini}
R_{\mu\nu\rho\sigma}k^\mu E_{\hat i}{}^\nu n^\rho E_{\hat j}{}^\sigma 
&=-S^{-1}k^A n^B (D_A D_B S)  \delta_{\hat i\hat j}
=\frac{f_2f_1'S'-f_1 f_2'S'+2 f_1 f_2 S''}{4f_1 f_2 ^2 S}\delta_{\hat i\hat j},
\end{align}
where the prime denotes the partial differentiation with respect to $r$. 
The trace of (\ref{Rili}) is tantamount to 
\begin{align}
\label{}
R_{\mu\nu}k^\mu k^\nu =-(n-2) S^{-1}k^A k^B D_A D_B S.
\end{align}
The divergence of 
the Riemann tensor component
$R_{\mu\nu\rho\sigma}k^\mu E_{\hat i}{}^\nu k^\rho E_{\hat j}{}^\sigma $ in the pseudo-spherical symmetric case therefore 
 implies the strong curvature singularity in the sense of \cite{clarke,krolak}.


\begin{thebibliography}{99}



\bibitem{Aghanim:2018eyx}
N.~Aghanim \textit{et al.} [Planck],
Astron. Astrophys. \textbf{641}, A6 (2020)
doi:10.1051/0004-6361/201833910
[arXiv:1807.06209 [astro-ph.CO]].


\bibitem{Caldwell:1999ew}
R.~Caldwell,
Phys.\ Lett.\ B \textbf{545}, 23-29 (2002)
doi:10.1016/S0370-2693(02)02589-3
[arXiv:astro-ph/9908168 [astro-ph]].


\bibitem{Caldwell:2003vq}
R.~R.~Caldwell, M.~Kamionkowski and N.~N.~Weinberg,
Phys.\ Rev.\ Lett.\  \textbf{91}, 071301 (2003)
doi:10.1103/PhysRevLett.91.071301
[arXiv:astro-ph/0302506 [astro-ph]].
\bibitem{Dabrowski:2003jm}
M.~P.~Dabrowski, T.~Stachowiak and M.~Szydlowski,
Phys.\ Rev.\ D \textbf{68}, 103519 (2003)
doi:10.1103/PhysRevD.68.103519
[arXiv:hep-th/0307128 [hep-th]].



\bibitem{Maeda:2018hqu} 
H.~Maeda and C.~Mart\'{\i}nez,
PTEP \textbf{2020}, no.4, 043E02 (2020)
doi:10.1093/ptep/ptaa009
[arXiv:1810.02487 [gr-qc]].
  
\bibitem{Hawking:1973uf} 
 S.~W.~Hawking and G.~F.~R.~Ellis,
{\it The Large scale structure of space-time},
(Cambridge University Press, Cambridge, 1973).
  
  
\bibitem{Penrose:1964wq}
R.~Penrose,
Phys.\ Rev.\ Lett.\  \textbf{14} (1965), 57-59
doi:10.1103/PhysRevLett.14.57


\bibitem{Hawking:1969sw}
S.~Hawking and R.~Penrose,
Proc.\ Roy.\ Soc.\ Lond.\ A \textbf{A314} (1970), 529-548
doi:10.1098/rspa.1970.0021


\bibitem{Schon:1979rg} 
  R.~Schoen and S.~T.~Yau,
  Commun.\ Math.\ Phys.\  {\bf 65}, 45 (1979)
  doi:10.1007/BF01940959
  
  \bibitem{SchonYau} 
  R.~Schoen and S.~T.~Yau, ``Positive Scalar Curvature and Minimal Hypersurface Singularities,''
arXiv:1704.05490 [math.DG]. 

  \bibitem{Witten:1981mf} 
  E.~Witten,
  Commun.\ Math.\ Phys.\  {\bf 80}, 381 (1981) 
  doi:10.1007/BF01208277
  
\bibitem{Hawking:1971vc}
S.~Hawking,
Commun.\ Math.\ Phys.\  \textbf{25} (1972), 152-166
doi:10.1007/BF01877517




\bibitem{Friedman:1993ty} 
  J.~L.~Friedman, K.~Schleich and D.~M.~Witt,
  Phys.\ Rev.\ Lett.\  {\bf 71}, 1486 (1993)
  Erratum: [Phys.\ Rev.\ Lett.\  {\bf 75}, 1872 (1995)]
  doi:10.1103/PhysRevLett.75.1872, 10.1103/PhysRevLett.71.1486
  [gr-qc/9305017].
  
  
\bibitem{Galloway}
   G. J. Galloway, 
   Class. Quant. Grav.  {\bf 12}, L99 (1995) 
  doi: 10.1088/0264-9381/12/10/002

\bibitem{Morris:1988cz}
M.~Morris and K.~Thorne,
Am. J. Phys. \textbf{56}, 395-412 (1988)
doi:10.1119/1.15620


\bibitem{Morris:1988tu}
M.~Morris, K.~Thorne and U.~Yurtsever,
Phys. Rev. Lett. \textbf{61}, 1446-1449 (1988)
doi:10.1103/PhysRevLett.61.1446


\bibitem{Ellis1973}
H.~G.~Ellis,
J. Math. Phys. \textbf{14}, 104-118 (1973)
doi:10.1063/1.1666161

\bibitem{Bronnikov1973}
K.~A.~Bronnikov,
Acta Phys. Polon. B \textbf{4}, 251-266 (1973).

  
\bibitem{Maldacena:2004rf}
J.~M.~Maldacena and L.~Maoz,
JHEP \textbf{02}, 053 (2004)
doi:10.1088/1126-6708/2004/02/053
[arXiv:hep-th/0401024 [hep-th]].


  \bibitem{Gao:2016bin} 
  P.~Gao, D.~L.~Jafferis and A.~C.~Wall,
  JHEP {\bf 1712}, 151 (2017)
  doi:10.1007/JHEP12(2017)151
  [arXiv:1608.05687 [hep-th]].
  
\bibitem{Maldacena:2017axo}
J.~Maldacena, D.~Stanford and Z.~Yang,
Fortsch.\ Phys.\  \textbf{65}, no.5, 1700034 (2017)
doi:10.1002/prop.201700034
[arXiv:1704.05333 [hep-th]].


\bibitem{Maldacena:2018lmt}
J.~Maldacena and X.~Qi,
``Eternal traversable wormhole,''
[arXiv:1804.00491 [hep-th]].


\bibitem{Fisher:1948yn} 
  I.~Z.~Fisher,
Zh. Eksp. Teor. Fiz. \textbf{18}, 636 (1948)
[arXiv:gr-qc/9911008 [gr-qc]].
  
  

%
\bibitem{Bergmann:1957zza}
O.~Bergmann and R.~Leipnik,
Phys. Rev. \textbf{107}, 1157-1161 (1957)
doi:10.1103/PhysRev.107.1157  
  
  
  %
\bibitem{Buchdahl:1959nk}
H.~A.~Buchdahl,
Phys. Rev. \textbf{115}, 1325-1328 (1959)
doi:10.1103/PhysRev.115.1325

  \bibitem{jnw1968}
A.I.~Janis, E.T.~Newman, and J.~Winicour, 
Phys. Rev. Lett., {\bf 20}, 878 (1968)
doi:10.1103/PhysRevLett.20.878

%
\bibitem{Wyman:1981bd}
M.~Wyman,
Phys. Rev. D \textbf{24}, 839-841 (1981)
doi:10.1103/PhysRevD.24.839



\bibitem{Gibbons:2003yj} 
 G.~W.~Gibbons,
``Phantom matter and the cosmological constant,''
[arXiv:hep-th/0302199 [hep-th]].
  
  \bibitem{Gibbons:2017jzk} 
  G.~W.~Gibbons and M.~S.~Volkov,
  JCAP {\bf 1705}, 039 (2017)
  doi:10.1088/1475-7516/2017/05/039
  [arXiv:1701.05533 [hep-th]].
    

\bibitem{wald}
R.M.~Wald, {\it General Relativity}, (University of Chicago Press,
1984).




\bibitem{Maeda:2016ddh}
H.~Maeda and C.~Mart\'{\i}nez,
Eur. Phys. J. C \textbf{78}, no.10, 860 (2018)
doi:10.1140/epjc/s10052-018-6334-7
[arXiv:1603.03436 [gr-qc]].

\bibitem{JNWhigher}
B. C. Xanthopoulos and T. Zannias, 
Phys. Rev. D {\bf 40}, 2564 (1989)
doi:10.1103/PhysRevD.40.2564.

\bibitem{Abdolrahimi:2009dc} 
  S. Abdolrahimi and A. A. Shoom,
  Phys.\ Rev.\ D {\bf 81}, 024035 (2010)
  doi:10.1103/PhysRevD.81.024035
  [arXiv:0911.5380 [gr-qc]].



\bibitem{Yilmaz}
  H. Yilmaz, 
Phys. Rev. {\bf 111}, 1417 (1958)
doi:10.1103/PhysRev.111.1417. 


\bibitem{Torii:2013xba}
T.~Torii and H.~Shinkai,
Phys. Rev. D \textbf{88}, 064027 (2013)
doi:10.1103/PhysRevD.88.064027
[arXiv:1309.2058 [gr-qc]].



\bibitem{Tabensky-Taub}
R. Tabensky and A. H. Taub, 
 Commun.Math. Phys. \textbf{29}, 61 (1973) 
 doi:10.1007/BF01661153
  
\bibitem{Singh}
T. Singh, 
Gen. Rel. Grav. \textbf{5}, 657 (1974) 
 doi:10.1007/BF00761923  
 
\bibitem{Buchdahl1978} 
H. A. Buchdahl, 
Gen. Rel. Grav. \textbf{9}, 59 (1978)
doi:10.1007/BF00772551

\bibitem{Vuille:2007ws}
C.~Vuille,
Gen. Rel. Grav. \textbf{39}, 621-632 (2007)
doi:10.1007/s10714-007-0411-9


\bibitem{Erices:2015xua}
C.~Erices and C.~Mart\'{\i}nez,
Phys. Rev. D \textbf{92}, no.4, 044051 (2015)
doi:10.1103/PhysRevD.92.044051
[arXiv:1504.06321 [gr-qc]].
  
  \bibitem{Maeda:2019tqs}
H.~Maeda and C.~Mart\'{\i}nez,
Class. Quant. Grav. \textbf{36}, no.18, 185017 (2019)
doi:10.1088/1361-6382/ab293a
[arXiv:1904.01658 [gr-qc]].  
  
  \bibitem{Majumdar:1947eu} 
  S.~D.~Majumdar,
  Phys.\ Rev.\  {\bf 72}, 390 (1947).
  doi:10.1103/PhysRev.72.390

  \bibitem{Papapetrou} 
  A. Papapetrou, 
   Proc.  Roy. Iri. Aca. Section A: Mathematical and Physical Sciences
{\bf 51} 191-204 (1945-1948)

\bibitem{Gibbons:1982fy}
G.~Gibbons and C.~Hull,
Phys. Lett. B \textbf{109}, 190-194 (1982)
doi:10.1016/0370-2693(82)90751-1





\bibitem{Coley:2005sq}
A.~Coley, S.~Hervik and N.~Pelavas,
Class. Quant. Grav. \textbf{23}, 3053-3074 (2006)
doi:10.1088/0264-9381/23/9/018
[arXiv:gr-qc/0509113 [gr-qc]].


\bibitem{Coley:2009tx}
A.~Coley, S.~Hervik and N.~Pelavas,
Class. Quant. Grav. \textbf{26}, 125011 (2009)
doi:10.1088/0264-9381/26/12/125011
[arXiv:0904.4877 [gr-qc]].


\bibitem{Kachru:2008yh}
S.~Kachru, X.~Liu and M.~Mulligan,
Phys. Rev. D \textbf{78}, 106005 (2008)
doi:10.1103/PhysRevD.78.106005
[arXiv:0808.1725 [hep-th]].




\bibitem{Arnowitt:1962hi}
R.~L.~Arnowitt, S.~Deser and C.~W.~Misner,
in \textit{Gravitation: An introduction to current research} (Chap. 7). Edited by Louis Witten. John Wiley \& Sons Inc., New York, London, 1962, pp. 227-265. Reprinted as
Gen. Rel. Grav. \textbf{40}, 1997-2027 (2008)
doi:10.1007/s10714-008-0661-1
[arXiv:gr-qc/0405109 [gr-qc]].

\bibitem{myersperry1986}
R.~C.~Myers and M.~J.~Perry,
Annals Phys. \textbf{172}, 304 (1986)
doi:10.1016/0003-4916(86)90186-7

\bibitem{Regge:1974zd} 
  T.~Regge and C.~Teitelboim,
  Annals Phys.\  {\bf 88}, 286 (1974) 
  doi:10.1016/0003-4916(74)90404-7

\bibitem{Henneaux:2006hk} 
  M.~Henneaux, C.~Mart\'{\i}nez, R.~Troncoso and J.~Zanelli,
  Annals Phys.\  {\bf 322}, 824 (2007)
  doi:10.1016/j.aop.2006.05.002
 [arXiv:hep-th/0603185 [hep-th]].
\bibitem{Saenz:2012ga} 
 S.~Garc\'{\i}a~S\'aenz and C.~Mart\'{\i}nez,
Phys.\ Rev.\ D {\bf 85}, 104047 (2012) 
 doi:10.1103/PhysRevD.85.104047
  [arXiv:1203.4776 [hep-th]].



\bibitem{Boonserm:2018orb}
P.~Boonserm, T.~Ngampitipan, A.~Simpson and M.~Visser,
Phys.\ Rev.\ D \textbf{98}, no.8, 084048 (2018)
doi:10.1103/PhysRevD.98.084048
[arXiv:1805.03781 [gr-qc]].

\bibitem{Yazadjiev:2017twg}
S.~Yazadjiev,
Phys. Rev. D \textbf{96}, no.4, 044045 (2017)
doi:10.1103/PhysRevD.96.044045
[arXiv:1707.03654 [gr-qc]].


\bibitem{Rogatko:2018smj}
M.~Rogatko,
Phys. Rev. D \textbf{97}, no.2, 024001 (2018)
doi:10.1103/PhysRevD.97.024001
[arXiv:1801.01987 [hep-th]].


\bibitem{Bhattacharya:2010zzb}
A.~Bhattacharya and A.~A.~Potapov,
Mod.\ Phys.\ Lett.\ A \textbf{25}, 2399-2409 (2010)
doi:10.1142/S0217732310033748


\bibitem{Abe:2010ap}
F.~Abe,
Astrophys.\ J.\  \textbf{725}, 787-793 (2010)
doi:10.1088/0004-637X/725/1/787
[arXiv:1009.6084 [astro-ph.CO]].


\bibitem{Tsukamoto:2016qro}
N.~Tsukamoto,
Phys.\ Rev.\ D \textbf{94}, no.12, 124001 (2016)
doi:10.1103/PhysRevD.94.124001
[arXiv:1607.07022 [gr-qc]].



\bibitem{Cremona:2018wkj}
F.~Cremona, F.~Pirotta and L.~Pizzocchero,
Gen.\ Rel.\ Grav.\  \textbf{51}, no.1, 19 (2019)
doi:10.1007/s10714-019-2501-x
[arXiv:1805.02602 [gr-qc]].

\bibitem{Roy:2019yrr}
P.~Dutta Roy, S.~Aneesh and S.~Kar,
Eur. Phys. J. C \textbf{80}, no.9, 850 (2020)
doi:10.1140/epjc/s10052-020-8409-5
[arXiv:1910.08746 [gr-qc]].

\bibitem{Hochberg:1998ha}
D.~Hochberg and M.~Visser,
Phys. Rev. D \textbf{58}, 044021 (1998)
doi:10.1103/PhysRevD.58.044021
[arXiv:gr-qc/9802046 [gr-qc]].



\bibitem{Kim:2013tsa}
S.~Kim,
J. Korean Phys. Soc. \textbf{63}, 1887-1891 (2013)
doi:10.3938/jkps.63.1887
[arXiv:1302.3337 [gr-qc]].


\bibitem{Misner:1964je}
C.~W.~Misner and D.~H.~Sharp,
Phys. Rev. \textbf{136}, B571-B576 (1964)
doi:10.1103/PhysRev.136.B571


\bibitem{Hayward:1994bu}
S.~A.~Hayward,
Phys. Rev. D \textbf{53}, 1938-1949 (1996)
doi:10.1103/PhysRevD.53.1938
[arXiv:gr-qc/9408002 [gr-qc]].





\bibitem{Bekenstein:1974sf}
J.~Bekenstein,
Annals Phys. \textbf{82}, 535-547 (1974)
doi:10.1016/0003-4916(74)90124-9


\bibitem{Maeda:1988ab}
K.~i.~Maeda,
Phys. Rev. D \textbf{39}, 3159 (1989)
doi:10.1103/PhysRevD.39.3159

\bibitem{Xanthopoulos:1992fm}
B.~Xanthopoulos and T.~Dialynas,
J. Math. Phys. \textbf{33}, 1463-1471 (1992)
doi:10.1063/1.529723


\bibitem{Klimcik:1993cia}
C.~Klim\v{c}\'{\i}k,
J. Math. Phys. \textbf{34}, 1914-1926 (1993)
doi:10.1063/1.530146


\bibitem{Bocharova:1970skc}
N.~Bocharova, K.~Bronnikov and V.~Melnikov,
Vestn. Mosk. Univ. Ser. III Fiz. Astron., no.6, 706-709 (1970)



\bibitem{Harada:2018ikn}
T.~Harada, B.~Carr and T.~Igata,
Class.\ Quant.\ Grav.\  \textbf{35}, no.10, 105011 (2018)
doi:10.1088/1361-6382/aab99f
[arXiv:1801.01966 [gr-qc]].


\bibitem{Blazquez-Salcedo:2018ipc}
J.~L.~Bl\'azquez-Salcedo, X.~Y.~Chew and J.~Kunz,
Phys. Rev. D \textbf{98}, no.4, 044035 (2018)
doi:10.1103/PhysRevD.98.044035
[arXiv:1806.03282 [gr-qc]].


\bibitem{paperII}
M. Nozawa, 
``Static spacetimes haunted by a phantom scalar field II: dilatonic charged solutions,'' 
to appear. 


\bibitem{paperIII}
M. Nozawa, 
``Static spacetimes haunted by a phantom scalar field III: asymptotically (A)dS solutions,'' 
to appear. 

\bibitem{Bogush:2020lkp} 
  I.~Bogush and D.~Gal'tsov,
  ``Generation of rotating solutions in Einstein-scalar gravity,''
  arXiv:2001.02936 [gr-qc].
  
  

\bibitem{Maeda:2007uu}
H.~Maeda and M.~Nozawa,
Phys. Rev. D \textbf{77}, 064031 (2008)
doi:10.1103/PhysRevD.77.064031
[arXiv:0709.1199 [hep-th]].


\bibitem{clarke}
C. J. S. Clarke and K. Kr\'olak,
J. Geom. Phys. {\bf 2} 127 (1985)
doi: 10.1016/0393-0440(85)90012-9

\bibitem{krolak}
K. Kr\'olak, J. Math. Phys.
{\bf 28}, 138 (1987)
doi: 10.1063/1.527795

\end{thebibliography}
\end{document}